\documentclass[pdflatex,sn-mathphys-ay]{sn-jnl}
\usepackage[utf8]{inputenc}

\usepackage{graphicx}%
\usepackage{multirow}%
\usepackage{amsmath,amssymb,amsfonts}%
\usepackage{amsthm}%
\usepackage{mathrsfs}%
\usepackage[title]{appendix}%
\usepackage{xcolor}%
\usepackage{textcomp}%
\usepackage{manyfoot}%
\usepackage{booktabs}%
\usepackage{algorithm}%
\usepackage{algorithmicx}%
\usepackage{algpseudocode}%
\usepackage{listings}%
\usepackage{nth}

\begin{document}

\newcommand{\R}{\mathbb{R}}
\newcommand{\Rn}{\mathbb{R}^n}
\newcommand{\PR}{\mathbb{P}}
\newcommand{\Rd}{{\displaystyle\mathbb{R}^2}}
\newcommand{\Rb}{\mathbb{\overline{R}}}
\newcommand{\Rbd}{{\displaystyle \mathbb{\overline{R}}^2}}
\newcommand{\Rbn}{{\displaystyle \mathbb{\overline{R}}^n}}
\newcommand{\I}{\mathbb{I}}
\newcommand{\Sv}{\mathbb{S}}
\newcommand{\Id}{{\displaystyle\mathbb{I}^2}}
\newcommand{\In}{{\displaystyle\mathbb{I}^n}}
\newcommand{\Z}{\mathbb{Z}}
\newcommand{\N}{\mathbb{N}}
\newcommand{\Hbb}{\mathbb{H}}
\newcommand{\lp}{\left(}
\newcommand{\rp}{\right)}
\newcommand{\lc}{\left[}
\newcommand{\rc}{\right]}
\newcommand{\lb}{\left\{}
\newcommand{\rb}{\right\}}
\newcommand{\lf}{\left.}
\newcommand{\ri}{\right.}
\newcommand{\id}{\stackrel{d}{=}}
\newcommand{\prob}[1]{\mathbb{P}\lp#1\rp}
\newcommand{\pr}[1]{\PR\lp#1\rp}
\newcommand{\esp}[2]{\mathbb{E}_#1\lc#2\rc}
\newcommand{\espe}[1]{\mathbb{E}\lc#1\rc}
\newcommand{\cb}{C\!\!\!\!/\!\!\!/\,}
\newcommand{\notprec}{\not\prec}
\providecommand{\abs}[1]{\left|#1\right|}
\newcommand{\casos}[4]{\lb\begin{array}{ll}#1,&#2\\#3,&#4\end{array}\ri}
\newcommand{\caso}[2]{\lb\begin{array}{l}#1\\#2\end{array}\ri}
\newcommand{\sign}{\textrm{sign}}
\newcommand{\corr}[2]{\textrm{Corr}\lp#1,#2\rp}
\newcommand{\cov}[2]{\textrm{Cov}\lp#1,#2\rp}
\newcommand{\Cov}[1]{\textrm{Cov}\lp#1\rp}
\newcommand{\var}[1]{\textrm{Var}\lp#1\rp}
\newcommand{\lcm}{\text{lcm}}
\setlength{\parindent}{0in}

\newcommand{\lrp}[1]{\left(#1\right)}
\newcommand{\lrc}[1]{\left[#1\right]}
\newcommand{\lrb}[1]{\left\{#1\right\}}
\newcommand{\E}[1]{\mathbb{E}\lc #1\rc}
\newcommand{\V}[1]{\mathbb{V}\mathrm{ar}\lc #1\rc}
\newcommand{\Es}[2]{\mathbb{E}_{#2}\lc #1\rc}
\newcommand{\Vs}[2]{\mathbb{V}\mathrm{ar}_{#2}\lc #1\rc}
\newcommand{\mse}[1]{\mathrm{MSE}\lrc{#1}}
\newcommand{\mise}[1]{\mathrm{MISE}\lrc{#1}}
\newcommand{\amise}[1]{\mathrm{AMISE}\lrc{#1}}
\newcommand{\pf}[2]{\frac{\partial #1}{\partial #2}}
\newcommand{\pftwo}[2]{\frac{\partial^2 #1}{\partial #2^2}}
\newcommand{\pfmix}[3]{\frac{\partial^2 #1}{\partial #2\partial #3}}
\newcommand{\norm}[1]{\left|\left| #1\right|\right|}
\newcommand{\tr}[1]{\text{tr}\left[#1\right]}
\newcommand{\inprod}[2]{\langle#1,#2\rangle}
\newcommand{\vlinel}[1]{\multicolumn{1}{|c}{#1}}
\newcommand{\vliner}[1]{\multicolumn{1}{c|}{#1}}

\newcommand{\Xcal}{\mathcal{X}}
\newcommand{\mcal}[1]{\mathcal{#1}}
\newcommand{\mat}[1]{\mathbf{#1}}
\newcommand{\ind}[1]{\mathbbm{1}_{\lrb{#1}}}
\newcommand{\rojo}[1]{{\color{red}#1}}
\newcommand{\verde}[1]{{\color{green}#1}}
\newcommand{\azul}[1]{{\color{blue}#1}}
\newcommand{\naranja}[1]{{\color{orange}#1}}

\title{Functional Regression Models with Functional Response: A New Approach and a Comparative Study}
\author*[1,2]{\fnm{Manuel}  \sur{Febrero--Bande}}\email{manuel.febrero@usc.es}
\author[3]{\fnm{Manuel} \sur{Oviedo--de la Fuente}}\email{manuel.oviedo@udc.es}
\author[1,4]{\fnm{Mohammad} \sur{Darbalaei}}\email{mohammad.darbalaei@rai.usc.es}
\author[4]{\fnm{Morteza} \sur{Amini}}\email{morteza.amini@ut.ac.ir}
\equalcont{These authors contributed equally to this work.}

\affil*[1]{\orgdiv{Dpt. of Statistics, Mathematical Analysis and Optimization}, \orgname{Univ. of Santiago de Compostela}, \orgaddress{\street{Facultade de Matem\'aticas, Campus Vida s/n}, \city{Santiago de Compostela}, \postcode{15782}, \state{A Coru{\~n}a}, \country{Spain}}}
\affil[2]{\orgname{Galician centre for Mathematical Research and Technology (CITMAga)},\orgaddress{\city{Santiago de Compostela},\country{Spain} }}

\affil[3]{\orgdiv{Dpt. of Mathematics, CITIC}, \orgname{Univ. de A Coru{\~n}a}, \orgaddress{\city{A Coru{\~n}a}, \country{Spain}}}

\affil[4]{\orgdiv{Dpt. of Statistics, School of Mathematics, Statistics and Computer Science}, \orgname{University of Tehran}, \orgaddress{\city{Tehran}, \country{Iran}}}

\abstract{This paper proposes a new nonlinear approach for additive functional regression with functional response based on kernel methods along with some slight reformulation and implementation of the linear regression and the spectral additive model. The latter methods have in common that the covariates and the response are represented in a basis and so, can only be applied when the response and the covariates belong to a Hilbert space, while the proposed method only uses the distances among data and thus can be applied to those situations where any of the covariates or the response is not Hilbert, typically normed or even metric spaces with a real vector structure. A comparison of these methods with other procedures readily available in \textsf{R} is perfomed in a simulaton study and in real datasets showing the results the advantages of the nonlinear proposals and the small loss of efficiency when the simulation scenario is truly linear. The comparison is done in the Hilbert case as it is the only scenario where all the procedures can be compared. Finally, the supplementary material provides a visualization tool for checking the linearity of the relationship between a single covariate and the response, another real data example and a link to a GitHub repository where the code and data is available.}

\keywords{Functional data analysis, Functional regression, Functional response,  Linear and Nonlinear models.}
\pacs[MSC Classification]{62R10, 62--04, 62G08, 62J02}

\maketitle

\section{Introduction}

Since the seminal book of \cite{Ramsay2005}, Functional Data Analysis (FDA) has become an important field in Statistics. The term FDA is reserved for discussing random functions, surfaces or volumes. In its primary definition, FDA deals with data consisting of curves with a common domain, typically a fixed interval $\mcal{S}=[a,b]$ ($\mcal{S}=[0,1]$ without loss of generality). These curves are usually assumed to belong to the Hilbert space $\mathcal{L}_2(\mcal{S}):=\{f:\mcal{S}\longrightarrow \R \textrm{ s.t. } \int_{\mcal{S}}|f(s)|^2ds<\infty\}$, equipped with an inner product, defined as: $\inprod{f}{g}:=\int_{\mcal{S}}f(s)g(s)ds$, its corresponding norm  $\|f\|_{2}:=\sqrt{\inprod{f}{f}}$  and the associated metric/distance among elements: $d_2(f,g):=\|f-g\|_{2}$. In practice, the curves are usually observed in a grid $a=t_1<t_2<\cdots<t_{r_S}=b\in \mcal{S}$ which might be dense or sparse, depending on measuring situations.

The key difference between FDA and multivariate data analysis (MV) is the dimension of the objects involved and hence, the dimension of the space. The observations in MV belong to Euclidean spaces, $\R^p$, whereas  FDA is usually reserved for talking about statistical objects like functions, images or surfaces, which are of infinite dimension. A couple of big differences can be drawn from  the infinite dimension of functional spaces. First, the objects of the space cannot be completely represented; thus, a dimension reduction tool must be performed when possible. Second, usual tricks in MV are now invalid in FDA due to the functional nature of the spaces. For instance, in a dense grid, the values of the curves at neighboring grid points become more and more interdependent, and an attempt to compute the inverse of the covariance matrix of the data become an ill-posed problem. The third major difference is that the common definitions of density and distribution function cannot be used for functional random elements due to the infinite dimension of the functional space. As a competitor of the MV, FDA has found its own path to solve these difficulties and has even evolved to deal with other kinds of functional spaces like Banach or metric (see, for instance, \cite{Ferraty2006}) or with more complex objects (see, for instance, \cite{Locantore1999}). A Banach space is a complete vector space equipped with a norm which does not necessarily have an inner product. A Hilbert space is always a Banach space with the associated  norm $\|f\|=\sqrt{\inprod{f}{f}}$. However, the
reverse is not true in general since it is impossible to define an inner product operator associated to a particular norm. Futhermore, every Banach space is a metric space to the associated metric $d(f,g):=\|f-g\|$ and, again, the reverse is not generally true. 

FDA has developed many techniques for functional data including outlier detection, clustering, classification, time series and regression. A Functional Regression Model (FRM) is a regression model in which the response and/or any covariates belong to a functional space.  The FRM models can be categorized depending on the type of the response and the covariates (functional or scalar), and the relationship between them (linear or nonlinear). The best-known functional regression problem in the literature is the functional linear regression with a scalar response,  sometimes referred to as the linear scalar-on-function model. In this model, the functional covariate $\mcal{X}$ is assumed to belong to $\mcal{L}_2(\mcal{S})$, and the linear relation is modeled by the usual integral linear operator $\int_{\mcal{S}}\mcal{X}(t)\beta(t)dt$, where $\beta\in \mcal{L}_2(\mcal{S})$ is an unknown coefficient function. The key idea for estimating this model is to represent $\beta$ and/or $\mcal{X}$ on an $\mcal{L}_2$ basis. This basis could be a fixed basis like Fourier or a B--spline basis (see, for instance, \cite[Chap. 5]{Ramsay2005}, \cite{Cardot2003}, \cite{Cardot2007a} or \cite{James2009}) or a  data--driven basis like functional Principal Components (fPC) or functional Partial Least Squares (fPLS) (see  \cite{Cardot1999},  \cite{Preda2005}, \cite{Reiss2007}, \cite{Hall2006}, \cite{Delaigle2012b}  or \cite{Febrero2017} among many others). In a nonlinear FRM, the relationship between the response and covariates is assumed to be an unknown general operator not necessarily linear. { Indeed, in these models, one can even relax the assumption that all the functional variables involved belong to a Hilbert space.} For the estimation of the unknown operator, several methods have been proposed in the literature, including the kernel approaches (see, for instance, \cite{Ferraty2006},  \cite{Aneiros2006}, \cite{Ferraty2009} or \cite{Febrero-Bande2013}) or by transforming the link among the covariate(s) and the response (see \cite{MullerYao2008}, \cite{Ferraty2013},  \cite{Chen2011}, \cite{McLean2014} or \cite{Fan2011}). The recent work by \cite{Rao2023} includes a list of nonlinear regression methods (Single Index, Multiple Index, Quadratic and Complex Quadratic) along with a new proposals based on neural networks with functional inputs.

The case when both the response and the covariate(s) are functional ({ also called} function-on-function regression in the literature) has attracted much less attention. For the linear case, see, for instance,  \cite[Chap. 16]{Ramsay2005}, \cite{Cuevas2002}, \cite{Chiou2003}, \cite{Ivanescu2014},  \cite{Chiou2016} or \cite{Beyaztas2020}. The nonlinear case is more scarce with \cite{MullerYao2008} being extended to functional response, \cite{Ferraty2012a} devoted to kernel approaches and \cite{Scheipl2015} and \cite{Qi2019}, where the chosen approach includes spline-based and fPC-based terms. We are not considering here those proposals that are treating the functional response case as a straightforward extension of the scalar response, just considering a loop of models along the grid points of the response (see, for instance, \cite{Rao2023, Rao2023b}). These proposals, generally speaking, are not taking into account the dependence among grid points specifically when the grid is dense nor the possible lack of continuity among models for consecutive grid points that are estimated independently. A discussion about this issue will be included later.  

This paper focuses on the yet scarce case of nonlinear functional regression models with functional responses that can accommodate  more than one functional covariate in the model. { We propose the kernel additive model (FKAM) that does not need to assume the Hilbert hypothesis and can estimate a nonlinear operator among spaces using only distances among trajectories. As a first advantage, this means that the kernel additive method can estimate regression models between Banach spaces where other types of distances among trajectories are available. Our proposal is based on \cite{Febrero-Bande2013} but using the {identity link}. As a second advantage,  to the best of our knowledge, all the available approaches in the literature focus on smooth models which is an additional assumption over the $\mcal{L}_2$. However, our proposal is the first publicly available implementation that can deal with non-smooth models in the scope of functional response. Non-smooth trajectories are not so rare. An example could be the pressure of a fluid along a pipeline that it is constructed splicing tubes of different section diameters or has new connections along its length. The measures before/after any junction may appear as non-smooth or discontinuous even though this signal belongs to the usual Hilbert space of functions. 

We also propose a review among methods that are readily available in \textsf{R} for functional response with our proposal. All these competitors are based on the representation of all functional information in bases, and so, are restricted to be applied when the response and the covariates are Hilbert. Specifically, we provide new implementations of linear and spectral additive models (\cite{MullerYao2008}) that allow a variety of basis expansions such as PC, PLS, Fourier or B--splines.}
For the spectral additive model (FSAM), our implementation uses a combination of smooth functions over the coefficients derived by a basis representation of the covariate and again, the flexibility of using several basis expansions is provided. These two methods rely on the hypothesis that all the covariates and the response belong to a Hilbert space as can be deduced from the existence of an appropriate basis for representing the trajectories.

In this paper, our proposal and different linear and nonlinear methods readily available in R are studied and compared with each other through several simulation scenarios and several real datasets. The examples will show the advantages of the nonlinear proposals and the small loss of efficiency when the scenario is truly linear. Furthermore, we present a graphical tool to assess the assumption of linearity or nonlinearity of the regression operator (in the supplementary material).

The structure of the paper is as follows. Section~\ref{sec:FRMFR} introduces the framework of functional regression models with functional responses.  Section~\ref{sec:FKAMFR} provides the proposed FKAM model. Section~\ref{sec:FLMFR} is devoted to the quite popular and competitive linear model. Section~\ref{sec:FNLMFR} gives an overview of the other developed approaches in FDA literature with implementations in R for estimating nonlinear operators. Section~\ref{sec:numstudies} provides the practical advice for applying these models and contains the simulation studies. Finally, two applications to real data are provided in Section~\ref{sec:realapp}. A graphical tool for assessing the assumption of linearity of the regression operator is presented in the supplementary material. The proposed methods are all included in a recently updated version of the \texttt{fda.usc} R package.

\section{Functional Regression Models with Functional Response}\label{sec:FRMFR}

Let $\{\mcal{Y},\{\mcal{X}^{1},\ldots,\mcal{X}^{J} \} \}$ be a vector of functional elements in the product space  $\mcal{F}_\mcal{Y}\times\mcal{F}_{\mcal{X}^1}\times\cdots\times\mcal{F}_{\mcal{X}^J}$, where each  $\mcal{F}_{\bullet}$ { is a Banach space} whose elements are measurable continuous functions almost everywhere, with domains $\{\mcal{S}_{\mcal{Y}}, \mcal{S}_1,\ldots, \mcal{S}_J\}$. { Examples of these spaces are, for instance, the \emph{Lebesgue Spaces}  i.e. $\mcal{F}_\bullet:=\{f:\mcal{S}_\bullet\rightarrow\R \textrm{ st } \int_{\mcal{S}_\bullet}|f(t)|^p dt<\infty\}$ (usually denoted by $\mcal{L}_p$)}, where $1\le p<\infty$ with the norm $\|f\|_p=\left(\int_{\mcal{S}_\bullet}|f(t)|^pdt\right)^{1/p}$ or $C(\mcal{S})$, the space of scalar functions on compact $\mcal{S}$ equipped with the max norm, $\|f\|_\infty=\max\{|f(s)|: s\in \mcal{S}\}$. { Recall that $\mcal{L}_2$ is the only one of the Lebesgue spaces that it is Hilbert and so, } possesses an inner product operator. 

The general case of a Functional Regression Model with Functional Response (FRMFR) is given by
\begin{equation}\label{FRMFR}
\mcal{Y}=\alpha+m(\mcal{X}^1,\cdots,\mcal{X}^J)+\epsilon,
\end{equation}
where $m:\mcal{F}_{\mcal{X}^1}\times\cdots\times\mcal{F}_{\mcal{X}^J}\longrightarrow \mcal{F}_\mcal{Y}$ is the regression operator among covariate spaces and the response space and  \{$\alpha, \epsilon\}\in \mcal{F}_\mcal{Y}$ with $\espe{\epsilon}=0$. For identifiability purposes and without loss of generality, we can assume that all covariates are centered. Thus, $\espe{m(\cdot)}=0$ and $\alpha:=\espe{\mcal{Y}}$. 
The second common hypothesis in regression models is that $\epsilon$ and the covariates { are uncorrelated,} i.e. $\cov{\mcal{X}^j(s)}{\epsilon(t)}=0 \; \forall s,t \in \mcal{S}_j\times\mcal{S}_{\mcal{Y}},\; \forall j=1,\ldots,J$.  

The task of identifying and estimating the operator $m$ in Model~\eqref{FRMFR} is quite challenging, not only due to the effect of the so--called course of dimensionality but also due to the complexity of the statistical objects involved. For instance, the mathematical instruments of a Hilbert space (inner product, orthonormal basis) make quite comfortable the treatment of the information stored in functions, making $\mcal{L}_2$ the most popular choice in the FDA literature by far.
 {But, the key to a successful regression model is the notion of distance or more specifically, the notion of neighborhood as the prediction for a new trajectory is constructed finding its neighbors in the sample/population and combining their responses. Neighborhood is defined in terms of the open balls of radius $\epsilon$, i.e. the data which is located at distance less than $\epsilon$: $B(\mcal{X}_0,\epsilon)=\{\mcal{X}: d(\mcal{X}_0,\mcal{X})<\epsilon\}$. So, if we restrict ourselves to a Hilbert space by the simplicity of the instruments, we are restricted to the neighborhood determined by the $\mcal{L}_2$ distance that, perhaps, is not the optimal for solving the statistical objective. Note that, a continuity property for the regression operator  is always needed, i.e. $\lim_{d(\mcal{X}^\prime,\mcal{X})\rightarrow 0} m(\mcal{X}^\prime)=m(\mcal{X})$, meaning that the choice of a particular functional space may affect which data is considered nearby and so, the construction of the regression operator $m$.}


In the following sections, we will consider different models for estimating the operator $m(\cdot)$ which provide different ways of predicting the functional response. The construction of these different models must take into account the characteristics of the functional spaces associated with the covariates and the response as well as the properties of the model itself.

\section{ Functional Kernel Additive Model}\label{sec:FKAMFR}

Our proposal is an extension of the kernel estimator presented in \cite{Febrero-Bande2013} for generalized models to the case of functional response { but using, of course, the identity link.} { Note that the usual definitions of exponential family distributions for scalar variates cannot be extended to an infinite dimensional response.} In this case, the  $m(\cdot)$ is conformed as a sum of the single effects of the covariates where each contribution is computed through kernel procedures. In the following, this model will be abbreviated by FKAMFR. 

Specifically, the model can be simply written as
\begin{equation}\label{FKAMFR}
	\mcal{Y}=\alpha+\sum_{j=1}^J\Phi_j(\mcal{X}^j)+\epsilon.
\end{equation}
where $\Phi_j$ is directly estimated using distances among data with kernel techniques. This means that any of the functional spaces associated with the response  ($\mcal{F}_\mcal{Y}$) or the covariates ($\mcal{F}_{\mcal{X}^j}$) could be a normed space (of course, including $\mcal{L}_2$ as a particular case). So, FKAMFR is the only alternative that is able to model relationships between functional spaces that, { at least one, is not Hilbert.} 

The case with a single functional covariate ($J=1$) was studied in \cite{Ferraty2012} where some convergence results were obtained. For $J>1$, there are only proposals for scalar responses. For instance, in \cite{Ferraty2009},  some boosting ideas were applied to estimate a functional kernel regression model with $J>1$ and some illustrations with $J=2$ were provided. Also, the aforementioned work by \cite{Febrero-Bande2013} extends the Backfitting algorithm (BF) for functional covariates in the context of generalized responses. Our proposal here follows the same line but for a functional response.

\paragraph{Backfitting for Functional Covariates and Response (BFFCFR)}

In short, the { backfitting (BF)} algorithm is an iterative procedure for estimating a regression model where the optimization of each covariate contribution is done assuming that the rest are fixed and performing a loop across covariates and iterations until convergence { \citep[See, e.g.,][]{Buja1989}}. So, $\textrm{BF}$ consists of two nested loops: the outer one is the overall iteration until convergence, and the inner one is the cycle across covariates. The BFFCFR algorithm mimics this structure as follows:

\begin{itemize}
	\item Initialize the estimators with $s=0$, $\hat{\alpha}^{(0)}:=\bar{\mcal{Y}}$ and $\hat{\Phi}_j^{(0)}:=0$ for $j=1,\ldots,J$.
	\item Construct  the fitted values as $\hat{\mcal{Y}}_i=\hat{\alpha}^{(s)}+\sum_{j=1}^J\hat{\Phi}_j^{(s)}(\mcal{X}_i^j)$.
	\item In the $s^{\rm th}$ iteration, update the current estimation of $\Phi_j$ for $j=1,\ldots,J$ by
	\begin{equation}\label{Bfopt}
		\widehat{\Phi}_j^{(s)}(\mcal{X}^j_0)=\frac{\sum_{i=1}^n\left(\mcal{Y}_i-\widehat{\mcal{Y}}_i^{(-j)}\right)K_j\left(\frac{d_j(\mcal{X}^j_0,\mcal{X}^j_{i})}{h_j}\right)}{\sum_{m=1}^n K_j\left(\frac{d_j(\mcal{X}^j_0,\mcal{X}^j_{m})}{h_j}\right)},
	\end{equation}
	where
	$$\widehat{\mcal{Y}}_i^{(-j)}=\hat{\alpha}^{(s)}+\sum_{k=1}^{j-1}\widehat{\Phi}_k^{(s)}(\mcal{X}^k_{i})+\sum_{k=j+1}^{J}\widehat{\Phi}_k^{(s-1)}(\mcal{X}^k_{i}),$$
	$K_j$ is a suitable kernel function, $d_j:\mcal{F}_{\mcal{X}^j}\times \mcal{F}_{\mcal{X}^j}\rightarrow \mathbb{R}^{+}$ is the metric associated with the norm of the functional space $\mcal{S}_{\mcal{X}^j}$ and $h_j>0$ is a proper bandwidth.
	\item Repeat iterations until convergence: $\|\Phi_j^{(s)}-\Phi_j^{(s-1)}\|<\delta$,  $\forall j=1,\ldots,J$. 
\end{itemize}

The final prediction after $s$ iterations is given by
$$
\widehat{\mcal{Y}}_0=\hat{\alpha}^{(s)}+\sum_{j=1}^J\widehat{\Phi}_j^{(s)}(\mcal{X}^{j}_0).
$$
The optimization step made in Model~\eqref{Bfopt} is computing a weighted mean of the pseudoresponses (the response minus the contribution of the other covariates) depending on the bandwidth parameter $h_j$. This parameter must be optimized by minimizing an appropriate loss function, typically related to the sum of the squared difference norms:  $\min_{\mathbf{h}} \sum_{i=1}^n\|\mcal{Y}_i-\hat{\mcal{Y}_i}\|_{\mcal{Y}}^2$. 
This regression model can be extended to metric spaces (with a real vector space structure) in which Model \eqref{Bfopt} can be computed. In any case, depending on the metrics/norms of the covariates, the optimization step can be more or less affordable although as $h_j$ is a scalar parameter, an exhaustive search along a grid is always possible. This grid can be defined as quantiles of the lower triangle of distances of covariates perhaps depending of the functions $K_j$. 


For the specific case when the functional response belongs to the Hilbert space, a possible alternative approach is to extend Model  \eqref{FKAMFR} by fitting Functional Kernel Additive Models with Scalar Response to each of the coefficients of an appropriate basis representation of the functional response (as it is implicitly done in Model \eqref{FLMFR}). It seems that there is no clear advantage to use the basis approximation instead of using directly the norms if the evaluation grid is dense. The choice between both must be evaluated among the two sources of numerical errors: the computation of the norms using the evaluation grid or the approximation of these norms through the basis representation. For instance, fitting Functional Kernel Additive Models with Scalar Response to each of the coefficients of an orthogonal basis representation of the functional response seems to be a better choice when we have sparse measurements of the response. Recall that, if a PC basis is used, $\|\mcal{Y}_i-\hat{\mcal{Y}}_i\|^2\approx\sum_{l=1}^L\lrp{y_i(l)-\hat{y}_i(l)}^2$ being $y_i(l)$ the coefficients of the PC representation and so, the dilemma lies in which side of the previous equation a greater numerical error is committed.    

{ The convergence of the algorithm (to one solution of the normal equations) can be guaranteed under the same assumptions respect to the smoothers (symmetric and shrinking) as in \cite[][Th. 9--10]{Buja1989}, at least for the $\mcal{L}_2$ case ensuring so, that  $\hat{\mcal{Y}}_0$ is a consistent estimator of $\mcal{Y}_0$. For other normed spaces, it is an open problem. More recently, \cite{Jeong2021} deduces existence and convergence rates,  using the Smooth Backfitting (SBF) algorithm for Hilbert response variables and covariates in Hilbert and semi--metric spaces or Riemannian manifolds. As in the multivariate framework, the use of SBF typically ensures the convergence of each $\hat{\Phi}_k$ to $\Phi_k$, whereas the $\textrm{BF}$ only ensures the global convergence of $\hat{\mcal{Y}}_0$ to $\mcal{Y}_0$ (see, again, \cite{Buja1989}). The superior properties of SBF are obtained at the cost of computing marginal densities for $(\mcal{X}^{(j)},\mcal{X}^{(k)})$ and $\mcal{X}^{(j)}$, which can be problematic for infinite--dimensional spaces. \cite{Jeong2021} solved this by imposing that $\mcal{X}^{(j)}$ must lie in a compact subset of $q_j$--dimensional Hilbert space or in a finite dimensional Riemannian manifold.}

\section{ Functional Linear Models}\label{sec:FLMFR}

The Functional Linear Model with Functional Response (FLMFR) { is defined among a Hilbert response $\mcal{Y}$ and Hilbert covariates $\{\mcal{X}_j\}_{j=1}^J$} as follows 

\begin{equation}\label{FLMFR}
	\mcal{Y}(t)=\alpha(t)+\sum_{j=1}^J\langle \mcal{X}^j,\beta_j \rangle(t)+\epsilon(t),
\end{equation}

where $\langle \mcal{X}^j,\beta_j \rangle(t):=\int_{\mcal{S}_j}\mcal{X}^j(s)\beta_j(s,t)ds$ with $\beta_j\in \mcal{F}_{\mcal{X}^j}\times\mcal{F}_\mcal{Y}$, $\alpha\in\mcal{F}_{\mcal{Y}}$ and $\epsilon\in\mcal{F}_\mcal{Y}$ is a zero mean process with covariance function $\cov{\epsilon(t^\prime)}{\epsilon(t)}=\Sigma_\epsilon(t^\prime,t)$.  Without loss of generality, all covariates and the response can be centered and then $\alpha\equiv 0$. As in the multivariate case, $\alpha$ can be estimated for non-centered variates as $ \hat{\alpha}=\bar{\mcal{Y}}-\sum_{j=1}^J\langle \bar{\mcal{X}}^j,\hat{\beta}_j \rangle$.

For the sake of simplicity, let us consider the linear model with just one covariate $\mcal{X}$ in the following. The key idea is to take advantage of the Hilbert structure for representing the response, the parameter $\beta$ and the covariate on an appropriate basis and approximate the Functional Regression problem by a Multivariate Linear Model (MLM) over the coefficients of the representation. See, for instance, chapter 16 of \cite{Ramsay2005}. This is a crucial step that helps to avoid some of the shortcomings that appear if we try to directly explain $\mcal{Y}(t)$ as a MLM over the values along the grid. Clearly, such MLM would be inefficient due to the collinearity among values in neighbor grid points, particularly, when the grid become denser. Indeed, the additional, and quite common in the literature, assumption that  $\cov{\epsilon(t^\prime)}{\epsilon(t)}=0\; \forall t^\prime\neq t$, { is only plausible} when the grid is sparse supposing that the dependence among consecutive points vanishes. { This additional assumption implies that, as the grid becomes denser ($\min |t^\prime-t|\rightarrow 0)$, the $\epsilon$ process is discontinuous almost everywhere and so, does not belong to $\mcal{L}_2$.} This assumption situates the functional regression model closer to the methods from Sparse Longitudinal Data Analysis literature than to FDA  { \citep[see, for instance,][]{Yao2005}.}  

Being explicit with the key idea, consider that we can approximately represent $\mcal{Y}$ in a basis $\theta$ with $L$ elements, $\mcal{X}$ in a basis $\eta$ with $K$ elements, and $\beta_1$ in a tensor product basis $(\eta^*,\theta^*)$ with $(K^*,L^*)$. { The usual choice is to let $\eta^*=\eta$, $\theta^*=\theta$, $K^*=K$ and $L^*=L$ to reduce computations when the bases are, for instance, orthogonal.} The functions are observed  in grids $t_1,\ldots,t_{r_T}\in \mcal{S}_{\mcal{Y}}$ and $s_1,\ldots,s_{r_S}\in \mcal{S}_{\mcal{X}^1}$, respectively. Then, the function evaluations over the grids and their approximate basis representations can be written in the following matrix forms:

\begin{eqnarray*}
	\mcal{Y}_i(t_j)&\approx&\sum_{l=1}^L y_{il}\theta_l(t_j), \; i=1,\ldots, n,\; j=1,\ldots,r_T \implies \mathbf{Y}\approx\mathbf{y}\boldsymbol{\theta} \\
	\mcal{X}_i(s_j)&\approx&\sum_{k=1}^K x_{ik}\eta_k(s_j), \; i=1,\ldots, n,\; j=1,\ldots,r_S  \implies \mathbf{X}\approx\mathbf{x}\boldsymbol{\eta} \\
	\beta(s_i,t_j)&\approx&\sum_{k=1}^{K^*}\sum_{l=1}^{L^*} b_{k,l}\eta_k^*(s)\theta_l^*(t)\; i=1,\ldots, r_S,\; j=1,\ldots,r_T \implies \boldsymbol{\beta}\approx\boldsymbol{\eta^*}^\top\mathbf{B}\boldsymbol{\theta^*} 
\end{eqnarray*}
where  $\mathbf{x}=(x_i(s_j))_{i,j=1}^{n,r_{S}}$, $\mathbf{y}=(y_i(t_j))_{i,j=1}^{n,r_{T}}$ are values at the grid points for $\mcal{X}$ and $\mcal{Y}$, respectively,  $\mathbf{B}=(b_{kl})_{k,l=1}^{K^*,L^*}$ is the coefficient matrix for $\beta$. Also,  $\boldsymbol{\eta}=(\eta_k(s_j))_{k,j=1}^{K,r_{S}}$, $\boldsymbol{\theta}=(\theta_l(t_j))_{l,j=1}^{L,r_{T}}$ (resp. $\boldsymbol{\eta^*}$, $\boldsymbol{\theta^*}$) are the evaluations of the tensor product basis on the double grid.

Similar to the multivariate case, Model~\eqref{FLMFR} can be estimated by minimizing the Residual Sum of Squares Norms $RSSN(\beta)=\sum_{i=1}^n\|\mcal{Y}_i-\hat{\mcal{Y}}_i\|^2$, which is represented by

\begin{equation}\label{RSSN}
	RSSN(\mathbf{B})\approx\sum_{i=1}^n\|\epsilon_i\|^2=\sum_{i=1}^n\|\mathbf{y}_i\boldsymbol{\theta}-\mathbf{x}_i\boldsymbol{\eta}\boldsymbol{\eta^*}^\top\mathbf{B}\boldsymbol{\theta^*}\|^2.
\end{equation}

Using this trick, the estimation of $\boldsymbol{\beta}$ relies on finding the $K^*\times L^*$ parameters of $\mathbf{B}$ instead of the $r_{S}\times r_T$ parameters of $\boldsymbol{\beta}$ (typically larger). Another advantage is that by choosing an appropriate basis (for instance, with orthogonal components), the dependence among columns of $\mathbf{X}$ vanishes making the estimation more stable. Therefore, the two basis must be chosen balancing their numerical properties and the quality of the representation. Note that the tensor product basis expansion of $\beta$ might use different basis functions  from those associated with $\mcal{X}$ and $\mcal{Y}$. As pointed out by \cite{Preda2011}, a PLS basis has an equivalence between the functional model and the Multiple Linear Model among representation coefficients in the basis for the covariate and the response. { This} result can be extrapolated to any type of basis. Also, suppose { that} some penalization on $\boldsymbol{\eta}$ or $\boldsymbol{\theta}$ (typically related to smoothing) is desired. In that case, an  appropriate penalization matrix and factor can be added to Eq.~\ref{RSSN}.  For an introductory view on this, see again, \cite{Ramsay2005}.

We extend the results of \cite{Febrero2017} to compute the conditional mean--square prediction error (CMSPE) of the scalar response for a new observation $\lrp{\mcal{Y}_0,\mcal{X}_0}$ by considering the special case of $K^*=K$, $L^*=L$ and employing the PC basis $\eta^*=\eta$ and $\theta^*=\theta$. Then, the following expression can be obtained for functional response case:
\begin{equation*}
	\espe{\norm{\mcal{Y}_0-\hat{\mcal{Y}}_0}^2|\boldsymbol{\mcal{X}}}=\sum_{l=1}^L\lrp{\hat{\sigma}_{(l)}^2+\frac{\hat{\sigma}_{(l)}^2}{n}\lrp{1+\sum_{k=1}^K\frac{x_{0,k}^2}{\hat{\lambda}_k}}}+\norm{R_K^{\mcal{X}}}^2+\norm{R_L^{\mcal{Y}}}^2,
\end{equation*}
where $\mcal{Y}_0=\sum_{l=1}^\infty y_{0,l}\theta_l$, $\hat{\mcal{Y}}_0=\sum_{l=1}^L\hat{y}_{0,l}\hat{\theta}_l$, $R_L^{\mcal{Y}}=\sum_{l=L+1}^\infty y_{0,l}\theta_l$, $R_K^{\mcal{X}}=\sum_{k=K+1}^\infty x_{0,k}\eta_k$, $\hat{\lambda}_k$ is the estimation of $k$-th eigenvalue for $\mcal{X}$, and $$\hat{\sigma}_{(l)}^2=\frac{1}{n-K}\sum_{i=1}^n\lrp{y_{il}-\sum_{k=1}^K\hat{b}_{kl}x_{ik}}^2.$$ 

Leaving aside the two remainder representation terms, $R_K^{\mcal{X}}$ and $R_L^{\mcal{Y}}$, which decrease as $K$ and $L$ increase, the role of these parameters is not symmetric. Although $L$ has an impact only in the remaining term to obtain a better approximation, $K$ can substantially affect the CMSPE depending on the values of the eigenvalues. Other effects of covariates in the CMSPE are related to $\hat{\sigma}^2_{(l)}$, which decreases as $K$ increases. So, the choice of $K$ seems more critical than the choice of $L$ and must be balanced its effect among  $\hat{\sigma}^2_{(l)}$ and $\hat{\lambda}_1,\ldots,\hat{\lambda}_K$.    

Different approaches found in the literature differ essentially in the type of basis employed, the type of penalization and the type of numerical integration method. For instance, \cite{Ivanescu2014} use the Penalized Functional Regression (PFR) presented by \cite{Goldsmith2011}, representing the covariates in a large number of eigenfunctions and the functional coefficient using a Penalized Spline Regression. 
The Linear Signal Compression approach (LSC) proposed by \cite{Luo2017} represents the $\beta$ coefficient using the eigenfunctions of the covariate as $\boldsymbol{\eta}$ and the eigenfunctions of the response as $\boldsymbol{\theta}$. So, $\beta$ is constructed as a tensor product basis expansion of the Karhunen--Loeve representations of $\mcal{X}$ and $\mcal{Y}$. 

The extension to multiple covariates of any of the above approaches is straightforward since the FLMFR is estimated through an approximation to a Multiple Linear Model among the coefficients of the representation of the covariates and the coefficients of the representation of the response. The challenge of all these approaches is how to balance the parsimony of the model with the quality of the representation for $\mcal{X}$'s  and $\mcal{Y}$.  

\section{ Other Functional Nonlinear Models}\label{sec:FNLMFR}
Other functional nonlinear regression models are described in the following two subsections.  The first one is based on \cite{MullerYao2008} and our contribution is a simpler implementation of { this procedure}. This model requires that the covariates and the response belong to a Hilbert space. The second subsection briefly describe other available nonlinear competitors with implementations in R. All of them are compared in the numerical study section.

\subsection{Functional Spectral Additive Model}
{One way to consider more complex relationships than those described by the linear model is to use the idea put forward by \cite{MullerYao2008}. Instead of assuming, as in a FLMFR, that the expected values of the fPC scores of the functional response are a linear function of the fPC scores of the covariates, we can relax this linear assumption to allow for additive nonlinear components. Since fPCs are related to the spectral decomposition of the covariance operators, we have added the adjective "Spectral" (which is not in the original) to the name of the procedure, although, of course, the model can also be constructed using other types of basis.}


Suppose that $\mcal{X}^j$ has the following truncated basis representation 
$$\mcal{X}^j\approx\sum_{k=1}^{K_j}x_k^j\eta_k.$$
For a scalar response, the Functional Spectral Additive Model (FSAM) is defined as

\begin{equation}\label{FSAM}
y=\alpha+\sum_{j=1}^J\Phi_j(\mcal{X}^j)+\epsilon=\alpha+\sum_{j=1}^J\sum_{k=1}^{K_j}f_{jk}(x_k^j)+\epsilon,
\end{equation}
where $f_{jk}$ is a general nonlinear function, $x_k^j$ is the $k^{\rm th}$ coefficient of the basis for representing $\mcal{X}^j$ and fulfilling, for identifiability purposes, that $\espe{f_{jk}(x_k^j)}=0\; \forall j,k$. Again, as before, $\epsilon$ is a zero mean vector with covariance operator $\Sigma_\epsilon$. 
FLM is the particular case of this model when $f_{jk}$ is linear. This additive model can only be applied to Hilbert covariates due to the use of basis representations. One of the advantages of  additive models in the multivariate framework is that the contribution of each covariate can be easily interpreted. This advantage is diluted in the functional framework because the interpretation of the effects of a functional covariate must be done jointly considering the functions  $f_{jk}$ along with the basis and the estimated coefficients. Except in some simple cases, this can be a challenging task. On the other hand, choosing an orthogonal basis, the collinearity problem is minimized among functions associated with the same covariate.

The extension to functional response (FSAMFR) has been done through fitting a FSAM (\ref{FSAM}) for each  coefficients $y_1,\ldots,y_L$ of the basis representation $\mcal{Y}\approx\sum_{l=1}^Ly_l\theta_l$,

\begin{equation}\label{FAMFRb}
y(l)=\alpha(l)+\sum_{j=1}^J\sum_{k=1}^{K_j}f_{jk}^{(l)}(x_k^j)+\epsilon(l),\; l=1,\ldots,L.
\end{equation}

If the representation of the response is done on an orthogonal basis, the equations in Model (\ref{FAMFRb}) deal with uncorrelated responses and thus, we can expect only a small (or null) dependence among $y(l)$ for $l=1,\ldots,L$. In this model, although there is no closed form for CMSPE, like in the linear case, it seems that their constructions are similar, the closed form should be similar to the linear case, but it may include some additional conditions over $f_{jk}^{(l)}$. In our experience, Model (\ref{FAMFRb}) has a certain overfitting tendency that can be controlled by imposing restrictions on the flexibility of $f_{jk}^{(l)}$. 
 
\subsection{Other Nonlinear Approaches}
Some other nonlinear methods for regression models with functional response are described in the FDA literature with certain similarities with the above methods. The approach by \cite{Scheipl2015}, called Functional Additive Mixed Model (FAMM), is similar to Model \eqref{FKAMFR} in the sense that an additive model is constructed for each grid point of the response being the covariates of different nature (scalar, functional, categorical, { etc.). Focusing} on functional covariates, the FAMM is constructed as 

\begin{equation*}\label{FAMM}
\mathcal{Y}(t)=\alpha(t)+\sum_{j=1}^J f_j(\mathcal{X}^{(j)},t)+\epsilon(t),
\end{equation*}
where $\epsilon(t)$ is an i.i.d. Gaussian noise with variance $\sigma_\epsilon^2$. This latter assumption is unreasonable for a functional framework in a Hilbert space with a dense grid. The reason is that such independent process $\epsilon$ along $t$ does not belong to a Hilbert space as it is discontinuous almost everywhere ($\lim_{t\rightarrow t_0}\epsilon(t)\ne\epsilon(t_0)$). This assumption is only acceptable when the grid is sparse and the gap among consecutive grid points is large enough to consider that the dependence has vanished. 

The $f_j$ function for functional effects can add a linear or nonlinear contribution to the model. In the linear case,  $f_j(\mcal{X}^{(j)},t):=\int_{S_j}\mcal{X}^{(j)}(s)\beta_j(s,t)ds$ (similar to Model \eqref{FLMFR}), which might be estimated using a B-spline basis for representing $\mcal{X}^{(j)}$ and $\beta_j$. In the nonlinear case, one might consider  $f_j(\mcal{X}^{(j)},t):=\int_{S_j}F_j\left(\mcal{X}^{(j)}(s),s,t\right)ds=\int_{S_j}F^j_{X,s}(x(s),s)F_t(t)ds$, which also can be treated using B-spline expansions for $F^j_{X,s}$ (bivariate) and $F_t$. Typically, a smoothness penalty term is added to the estimation of the model to avoid, particularly in the latter case, the inflation in the number of parameters. 

An alternative idea for the nonlinear approach is provided by  \cite{Qi2019} who assumed $$F_j\left(\mcal{X}^{(j)}(s),s,t\right)=\sum_{k=1}^\infty G_k(\mcal{X}^{(j)}(s),s)\phi_k(t),$$ which is called \emph{Decomposition Induced by the Signal Compression} (DISC). The goal of this proposal is to find the partial $\sum_{k=1}^K G_k(\mcal{X}^{(j)}(s),s)\phi_k(t)$ with the minimum prediction error among all equivalent expansions of $F_j$. This can be done in two steps: (i) obtaining $\hat{G}_k$ by solving a generalized eigenvalue problem with a smoothness penalty derived from a Sobolev norm, and (ii) estimating $\phi_k$ given $\hat{G}_k$. Also, in practice $\hat{G}_k$, is estimated using a tensor product B-spline basis, which induces more smoothness in the final estimator. 

The main drawback of these approaches, apart from the high number of parameters, is how to generate the basis expansion of $G_k$ at  $\mcal{X}^{(j)}$. The tensor product B-spline basis associated with the values $\mcal{X}^{(j)}(s)$ must be constructed using the training sample. The values of these bases for new (future) curves may exceed the interval of the training sample, and this may produce a wiggly prediction for this new observation. Besides, if there is an outlier in the training sample, the range for the B-spline expansion could be distorted. In such a situation, the location of the knots (and so, the elements of the basis associated with them) may be  uninformative, which results in a considerable bias of the final estimator. However, there is no problem with the basis expansion at points $s$ or $t$ since the intervals of these values are fixed.

\section{Numerical Studies}\label{sec:numstudies}

This section compares the methods for functional response that are readily available in R. Our implementations, FLMFR, FSAMFR, and FKAMFR, are available in the {\tt fda.usc} package through the commands {\tt fregre.lm.fr}, {\tt fregre.sam.fr} and {\tt fregre.kam.fr} and are compared with the four mentioned competitor methods (namely, PFR, FAMM, LSC, and DISC). PFR and FAMM methods are available in the {\tt refund} package through the command {\tt pffr}, where the argument {\tt formula} allows us to include linear {\tt ffpc, ff} or nonlinear terms {\tt sff}. The authors considered the latter as an experimental feature. LSC and DISC methods are available in the {\tt FRegSigCom} package through the commands {\tt cv.sigcom} and {\tt cv.nonlinear}. However, this package is not currently maintained; its latest version was published in November 2018. Note that in this comparison, FLMFR, PFR, and LSC are linear, while FSAMFR, FKAMFR, FAMM and DISC are nonlinear methods. In all estimation cases when a basis is needed, we have chosen a principal component basis for each covariate or response with a length that ensures the percentage of variability explained (PVE) by the basis is over 95\%.

For each simulation scenario, a training sample $(\vec{\mcal{X}}_i, \mcal{Y}_i)_{i=1}^n$ with size $n\in \{100,200\}$ is generated, which is used for training all models. A prediction (validation) set $(\vec{\mcal{X}}_i^{p}, \mcal{Y}_i^p)_{i=1}^{n_p}$ with size $n_p=100$  is also generated to examine the quality of the estimates out of the training sample. The simulation scenarios will be generated by fixing $\rho^2$ defined as  
$$\rho^2=1-\frac{\espe{\|\epsilon\|^2}}{\espe{\|\mcal{Y}-\mu_{\mcal{Y}}\|^2}}.$$

Therefore, the methods will be compared using empirical measures of it
 
$$
R_{{\rm e}}^2=1-\frac{\sum_{i=1}^{n}\|\mcal{Y}_i-\hat{\mcal{Y}}_i\|^2 }{\sum_{i=1}^{n}\|\mcal{Y}_i-\bar{\mcal{Y}}\|^2},
$$
$$
R_{{\rm p}}^2=1-\frac{\sum_{i=1}^{n_{p}}\|\mcal{Y}_i^{p}-\hat{ \mcal{Y}}_i^{p}\|^2}{\sum_{i=1}^{n_{p}}\|\mcal{Y}_i^{p}-\bar{\mcal{Y}}\|^2},
$$
where $\hat{\mcal{Y}}_i$, $\hat{\mcal{Y}}_i^p$ are the estimates over the training and the validation set, respectively.

\subsection{Simulation Study}

Eight scenarios were constructed to compare methods: four linear and four nonlinear. In order to compare all procedures, the eight scenarios are Hilbertian, i.e. the response and the covariates belong to a Hilbert space of functions. In four scenarios the response depends only on the first covariate and in the rest, the response is generated using the two functional covariates. The first covariate ($\mcal{X}^{(1)}$) is generated from a zero-mean Ornstein-Uhlenbeck process in $[0,1]$, with the following covariance 
function 
$$\Sigma_1(u,v)=\frac{\sigma^2}{2\theta_1}\exp\bigg\{-\theta_1(u+v)\bigg\}\bigg(\exp\big\{2\theta_1\min(u,v)\big\}-1\bigg).$$ 
Here, we let $\sigma=1$ and $\theta_1=0.2$. 
The second covariate ($\mcal{X}^{(2)}$) is generated from a zero-mean Gaussian process with the following covariance function
\begin{equation}\label{expo}
\Sigma_2(u,v)=\sigma^2\exp\left(\frac{-|u-v|}{\theta_2}\right).
\end{equation}
In this case, we set $\sigma^2=0.5$ and $\theta_2=0.7$. Both covariates were generated on an equi-spaced grid of $51$ points.  The rule $\textrm{PVE}>95\%$ applied to the covariates leads to select four PCs for $\mcal{X}^{(1)}$ and five for $\mcal{X}^{(2)}$. 

We used the following four scenarios to generate the functional response.
\begin{itemize}
    \item Linear smooth (LS): The regression model is 
$$\mcal{Y}=\Delta\left(\inprod{\mcal{X}^{(1)}}{\beta_1^{\rm LS}}+\inprod{\mcal{X}^{(2)}}{\beta_2^{\rm LS}}\right)+\epsilon,$$ 
with constant $\Delta$ and the following coefficient functions
$$\beta_1^{\rm LS}(u,v)=6\sqrt{uv}\sin(4\pi v)$$
 and
$$\beta_2^{\rm LS}(u,v)=-(uv+1)\cos(2\pi \sqrt{uv}). $$
 
   \item Linear non-smooth (LNS): The regression model is 
$$\mcal{Y}=\Delta\left(\inprod{\mcal{X}^{(1)}}{\beta_1^{\rm LNS}}+\inprod{\mcal{X}^{(2)}}{\beta_2^{\rm LNS}}\right)+\epsilon,$$ 
with the following coefficient functions 
 $$\beta_1^{\rm LNS}(u,v)=\left\{\begin{array}{ll}
    5.6\exp(u^3)\left(\frac{v-0.1}{0.25}\right)^3 & v\in[0,\frac{1}{3}) \\ 6.3\left(\frac{v-0.6}{0.25}\right)^5 & v\in[\frac{1}{3},\frac{3}{4}) \\ 
    -28\left(\frac{v-1}{0.5}\right)^2\cos(\frac{\pi}{2}u) &  v\in[\frac{3}{4},1] \end{array}\right.$$ and $$\beta_2^{\rm LNS}(u,v)=\left\{\begin{array}{ll} -20v^2\cos(2\pi(2u-1)(2v-1)) & v\in[0,\frac{1}{2}) \\ 
    (2-3v)^2 & v\in[\frac{1}{2},1] \end{array}\right. $$

\item Nonlinear smooth (NLS):  The regression model is 

$$\mcal{Y}=\Delta\left(\inprod{\Phi_1(\mcal{X}^{(1)})}{\beta_1^{\rm LS}}+\inprod{\Phi_2(\mcal{X}^{(2)})}{\beta_2^{\rm LS}}\right)+\epsilon,$$ 
with  the following nonlinear operators 
$$\Phi_1(\mcal{X})(s)=\exp\left\{\frac{1}{2}(1-\mcal{X}(s)^2)\right\}$$ 
and 
$$\Phi_2(\mcal{X})(s)=1+\frac{\mcal{X}(s)^2}{5}+\cos(2\pi(\mcal{X}(s)/{2}-1))\sin(2\pi(\mcal{X}(s)/{2}-1)).$$

\item Nonlinear non--smooth (NLNS): The regression model is 
$$\mcal{Y}=\Delta\left(\inprod{\Phi_1(\mcal{X}^{(1)})}{\beta_1^{\rm LNS}}+\inprod{\Phi_2(\mcal{X}^{(2)})}{\beta_2^{\rm LNS}}\right)+\epsilon.$$
\end{itemize}
 The only difference of the alternative four scenarios with just one covariate is that we only include the first term (with $\mcal{X}^{(1)})$ in the above models.  
 In all cases,  the error process $\epsilon$ is generated on an equispaced grid of $71$ points in $[0,1]$ from a zero-mean Gaussian process with a covariance operator available by Equation \eqref{expo},  where $\sigma^2=0.5$, $\theta=0.3$, and $\Delta=1/71$. In order to control the signal-to-noise ratio, the $\epsilon$ process is re-scaled to set $\rho^2=0.8$.  To achieve this correlation, each error trajectory is multiplied by $$C_\rho=\sqrt{\frac{1-\rho^2}{\rho^2}\frac{\sum_i\|M_i-\bar{M}\|^2}{\sum_i\|\epsilon_i\|^2}},$$ where $M_i:=m(\mcal{X}_{i}^{(1)},\mcal{X}_{i}^{(2)})$ is the signal (conditional expectation given covariates). 

\begin{figure}
    \centering
\includegraphics[scale=0.5]{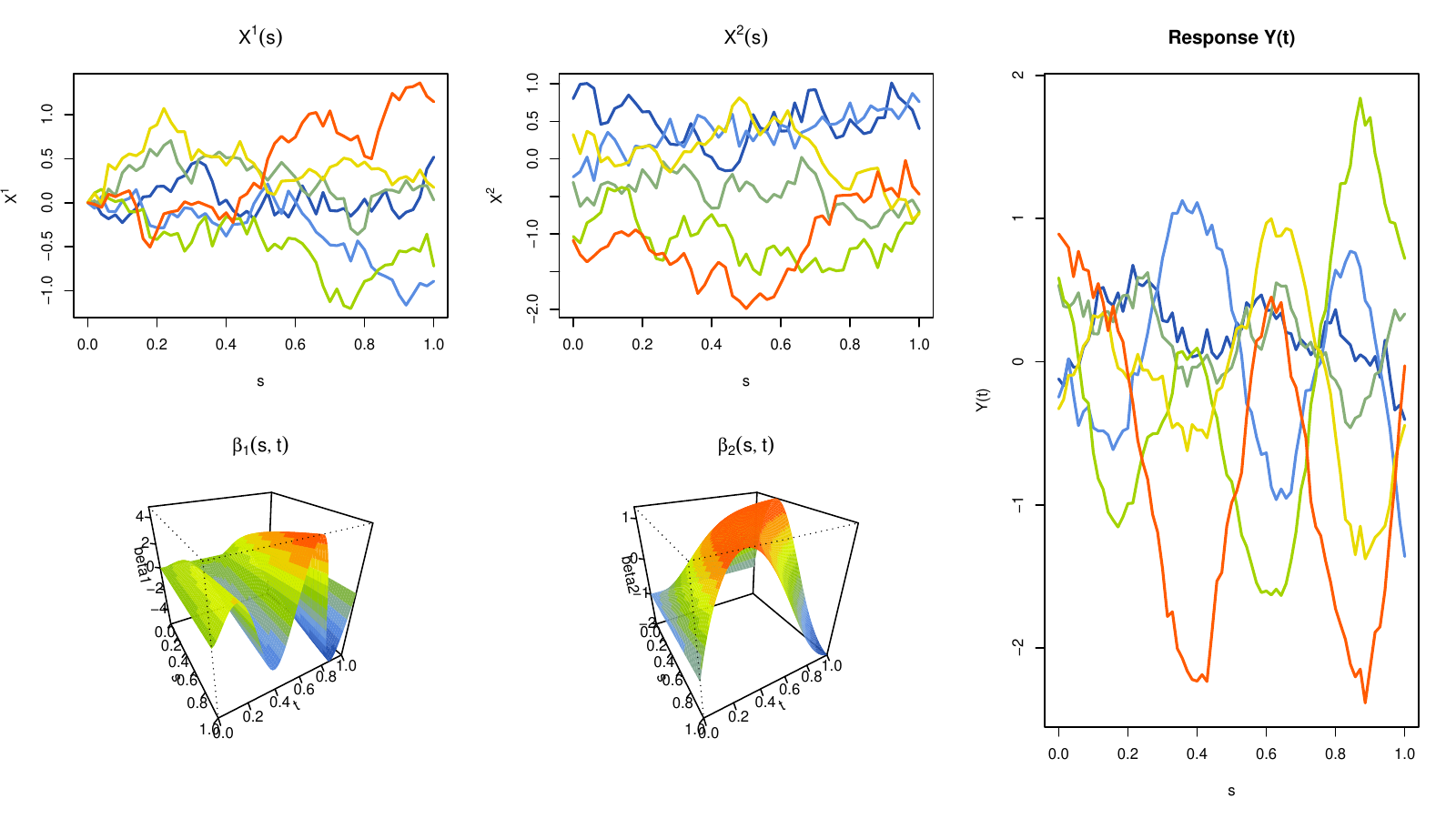}
    \caption{A sample of six realizations of the covariates and its responses jointly with the parameters for scenario 1 (Linear smooth).}
    \label{fig:scen1}
\end{figure}

\begin{figure}
    \centering
\includegraphics[scale=0.5]{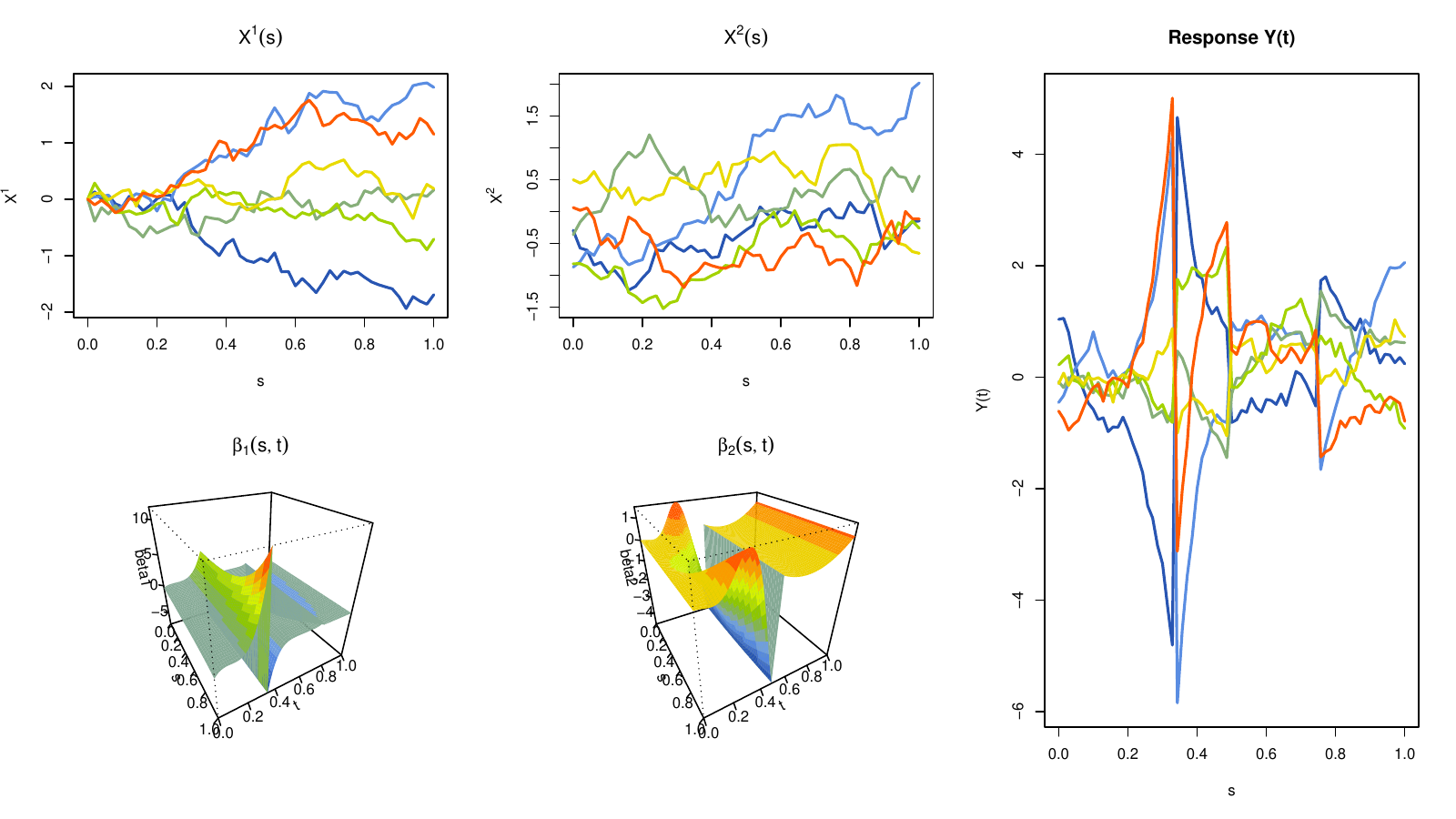}
    \caption{A sample of six realizations of the covariates and its responses jointly with the parameters for scenario 2 (Linear non--smooth).}
    \label{fig:scen2}
\end{figure}

\begin{figure}
    \centering
\includegraphics[scale=0.5]{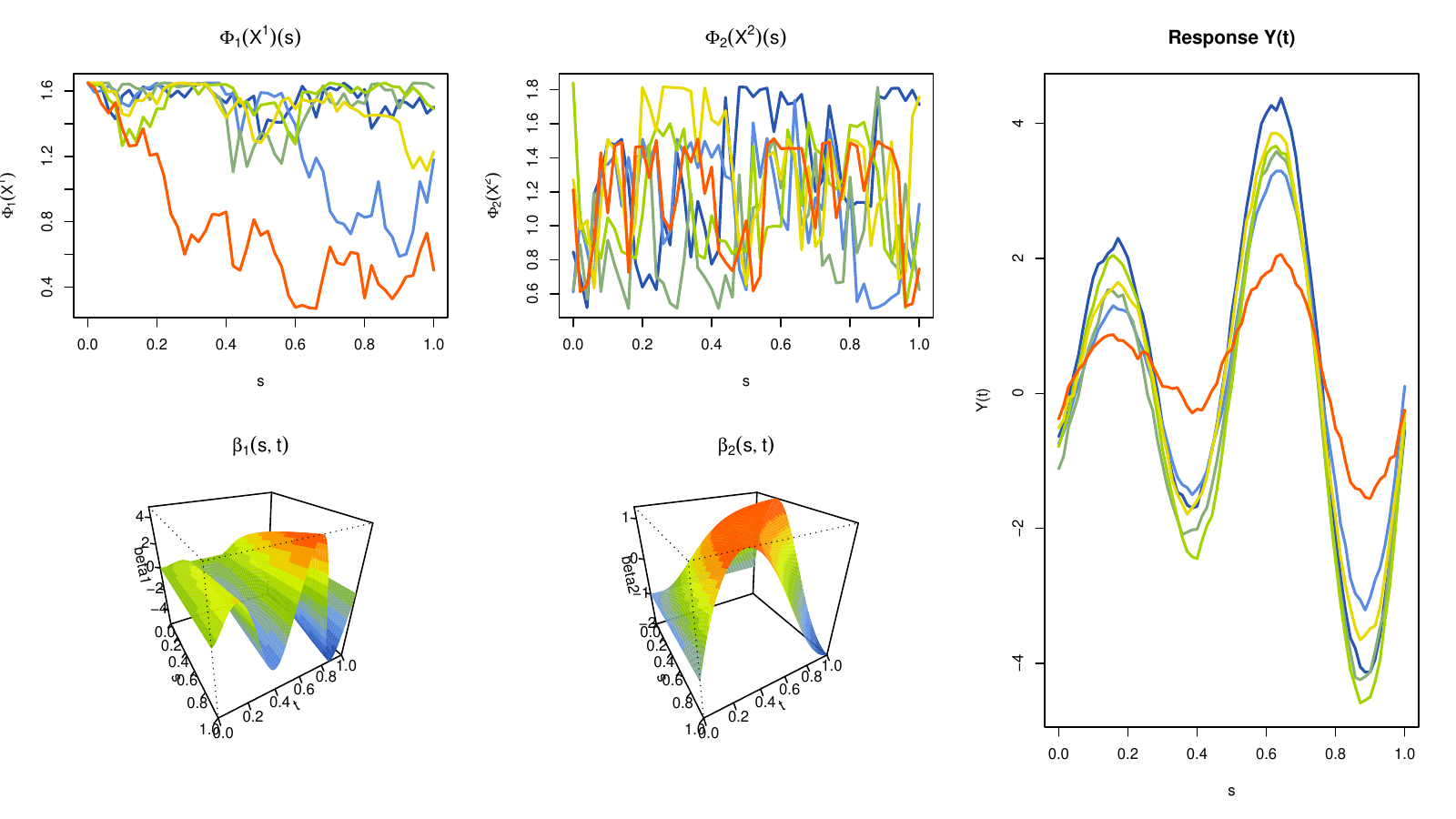}
    \caption{A sample of six realizations of the transformations of the covariates and its response jointly with the parameters for scenario 3 (Nonlinear smooth).}
    \label{fig:scen3}
\end{figure}

\begin{figure}
    \centering
\includegraphics[scale=0.5]{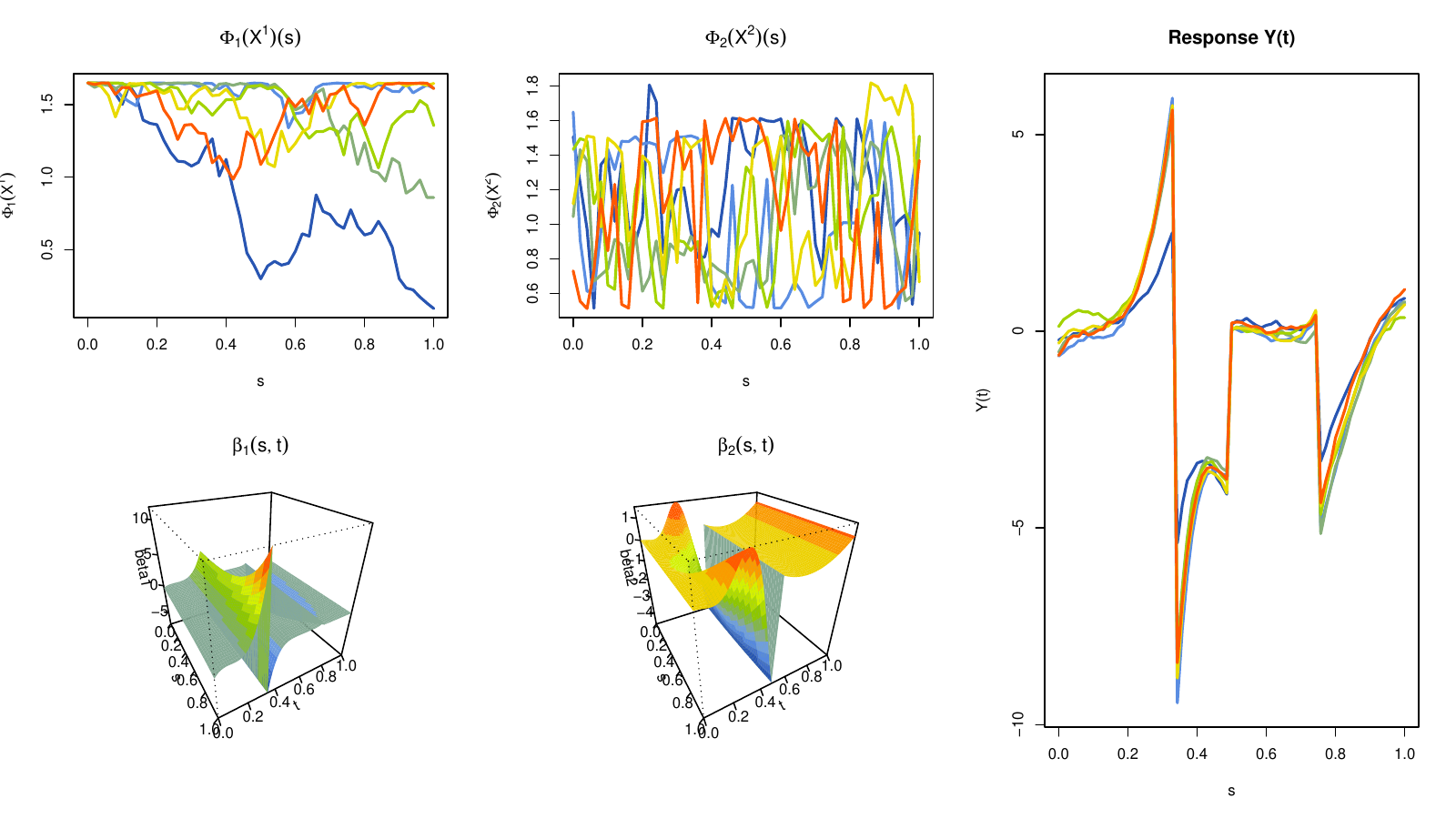}
    \caption{A sample of six realizations of the transformations of the covariates and the response jointly with the parameters for scenario 4 (Nonlinear non--smooth).}
    \label{fig:scen4}
\end{figure}

Scenarios LS (Figure \ref{fig:scen1}) and NLS (Figure \ref{fig:scen3}) are considered ``smooth'' in the sense that all the parameters and transformations involved are fairly smooth. In constrast, scenarios LNS (Figure \ref{fig:scen2}) and NLNS (Figure \ref{fig:scen4}) use some parameters that are discontinuous in some null-measure subsets of their domains. It seems quite ``non--smooth'' for the methods based on representing all the involved functions on a smooth basis like B--splines.

 \begin{table}
	\centering\small
\scalebox{1}[0.9]{
\begin{tabular}{ccccccccc}
\hline\hline
  &&& \multicolumn{3}{c}{Linear smooth}&&\\   \hline 
    \multicolumn{2}{c}{$n=100$}\\ \hline 
 & cov. & FLMFR & FSAMFR & FKAMFR & PFR & FAMM & LSC & DISC \\ 
  \hline
 $R_e^2$ & 1 & 0.818 & 0.854 & 0.845 & 0.811 & 0.275 & 0.806 & 0.806 \\ 
 $R_p^2$ & 1 & 0.781 & 0.741 & 0.768 & 0.776 & 0.188 & 0.793 & 0.793 \\ 
 $R_e^2$ & 2 & 0.818 & 0.857 & 0.844 & 0.812 & 0.321 & 0.810 & 0.807 \\ 
 $R_p^2$ & 2 & 0.781 & 0.735 & 0.750 & 0.776 & 0.223 & 0.787 & 0.782 \\ \hline
   \multicolumn{2}{c}{$n=200$}\\ \hline
 $R_e^2$ & 1 & 0.808 & 0.826 & 0.838 & 0.802 & 0.258 & 0.802 & 0.802 \\ 
  $R_p^2$ & 1 & 0.793 & 0.780 & 0.781 & 0.787 & 0.222 & 0.799 & 0.799 \\ 
  $R_e^2$ & 2 & 0.809 & 0.826 & 0.837 & 0.803 & 0.302 & 0.805 & 0.804 \\ 
  $R_p^2$ & 2 & 0.792 & 0.779 & 0.768 & 0.787 & 0.264 & 0.794 & 0.791 \\ 
   \hline\hline
  &&& \multicolumn{3}{c}{Linear non--smooth}&&\\ \hline 
    \multicolumn{2}{c}{$n=100$}\\ \hline 
 & cov. & FLMFR & FSAMFR & FKAMFR & PFR & FAMM & LSC & DISC \\ 
  \hline
 $R_e^2$ & 1 & 0.810 & 0.846 & 0.848 & 0.236 & 0.116 & 0.707 & 0.702 \\
 $R_p^2$ & 1 & 0.780 & 0.742 & 0.764 & 0.218 & 0.052 & 0.696 & 0.691 \\
 $R_e^2$ & 2 & 0.810 & 0.839 & 0.844 & 0.348 & 0.212 & 0.733 & 0.726 \\
 $R_p^2$ & 2 & 0.783 & 0.753 & 0.743 & 0.322 & 0.141 & 0.717 & 0.713 \\  \hline
   \multicolumn{2}{c}{$n=200$}\\ \hline
  $R_e^2$ & 1 & 0.802 & 0.818 & 0.841 & 0.232 & 0.102 & 0.705 & 0.705 \\
  $R_p^2$ & 1 & 0.782 & 0.770 & 0.772 & 0.223 & 0.073 & 0.696 & 0.695 \\ 
  $R_e^2$ & 2 & 0.801 & 0.816 & 0.839 & 0.341 & 0.195 & 0.729 & 0.726 \\
  $R_p^2$ & 2 & 0.786 & 0.776 & 0.758 & 0.333 & 0.163 & 0.720 & 0.718 \\ 
\hline\hline
   &&& \multicolumn{3}{c}{Nonlinear smooth}&&\\ \hline 
  \multicolumn{2}{c}{$n=100$}\\ \hline 
 & cov. & FLMFR & FSAMFR & FKAMFR & PFR & FAMM & LSC & DISC \\ 
  \hline
 $R_e^2$ & 1 & 0.099 & 0.841 & 0.867 & 0.096 & 0.279 & 0.055 & 0.808 \\
 $R_p^2$ & 1 & -0.089 & 0.727 & 0.725 & -0.077 & 0.168 & -0.049 & 0.781 \\
 $R_e^2$ & 2 & 0.098 & 0.825 & 0.847 & 0.096 & 0.292 & 0.052 & 0.781 \\
 $R_p^2$ & 2 & -0.105 & 0.685 & 0.693 & -0.090 & 0.152 & -0.052 & 0.748 \\  \hline
   \multicolumn{2}{c}{$n=200$}\\ \hline
  $R_e^2$ & 1 & 0.051 & 0.817 & 0.858 & 0.050 & 0.260 & 0.031 & 0.805 \\
  $R_p^2$ & 1 & -0.058 & 0.763 & 0.751 & -0.051 & 0.213 & -0.033 & 0.788 \\ 
  $R_e^2$ & 2 & 0.049 & 0.795 & 0.838 & 0.049 & 0.268 & 0.027 & 0.779 \\
  $R_p^2$ & 2  & -0.043 & 0.736 & 0.723 & -0.037 & 0.208 & -0.023 & 0.761 \\ 
   \hline\hline
   &&& \multicolumn{3}{c}{Nonlinear non--smooth}&&\\ \hline 
   \multicolumn{2}{c}{$n=100$}\\ \hline 
 & cov. & FLMFR & FSAMFR & FKAMFR & PFR & FAMM & LSC & DISC \\ 
  \hline
 $R_e^2$ & 1 & 0.102 & 0.831 & 0.873 & -10.717 & -10.656 & -3.995 & -3.358 \\
 $R_p^2$ & 1 & -0.099 & 0.703 & 0.722 & -10.268 & -10.205 & -3.919 & -3.194 \\
 $R_e^2$ & 2 & 0.095 & 0.789 & 0.810 & -9.611 & -9.538 & -3.573 & -2.942 \\
 $R_p^2$ & 2 & -0.099 & 0.623 & 0.624 & -9.642 & -9.574 & -3.661 & -3.015 \\  \hline
   \multicolumn{2}{c}{$n=200$}\\ \hline
  $R_e^2$ & 1 & 0.051 & 0.803 & 0.858 & -10.393 & -10.314 & -3.901 & -3.225 \\
  $R_p^2$ & 1 & -0.056 & 0.733 & 0.740 & -10.387 & -10.310 & -3.944 & -3.241 \\ 
  $R_e^2$ & 2 & 0.051 & 0.754 & 0.808 & -9.495 & -9.408 & -3.546 & -2.877 \\
  $R_p^2$ & 2 & -0.051 & 0.681 & 0.667 & -9.342 & -9.256 & -3.534 & -2.870 \\  
   \hline\hline
\end{tabular}
}

\caption{Simulation results ($R_e^2$ and $R_p^2$) for scenarios 1-4.}
\label{tab:scen1}
\end{table}

Table \ref{tab:scen1} shows the results for each scenario, with two samples sizes ($n\in\{100, 200\}$) and with one or two covariates (column \textrm{cov.}). The rows beginning with $R^2_e$ correspond to the estimation set (each cell is the average of $n\times 100$ values) and those starting with $R^2_p$ correspond to the prediction set (each cell is the mean of $n_p\times 100=10,000$ values).

The results for the Linear smooth scenario are expected  for all methods except FAMM even though we have tried its application using different options. The table shows the results for \texttt{splinepars=list(bs="ps", k=c(31,11), m=2))}, which seems to be flexible enough. The other methods obtain values close to the nominal level of $\rho^2=0.8$ with a slight tendency of FSAMFR and FKAMFR to overfitting that produces a certain underestimation in the prediction set. This behavior can be motivated by two reasons: First, the estimation with these two methods can be excessively flexible and some restrictions should be applied and  second, by a kind of border effect that in functional variates corresponds with trajectories in the prediction set far from the center. By this when the sample size is incremented, the effect is lowered. The winners in this scenario are the LSC and DISC methods closely followed by PFR and FLMFR taking into account the estimation and the prediction results. 

The results for Linear non-smooth scenario are clearly poor for PFR, and not so favorable for LSC and DISC. Recall that these methods are strongly based on representing the intermediate functions in a B--spline basis and the $\beta$ parameters of this scenario have discontinuities for certain lines (in any case, belonging to $\mcal{L}_2(S)\times \mcal{L}_2(T)$). The FLMFR method using PC basis is clearly the best one in this scenario although FSAMFR and FKAMFR are quite close.

In the nonlinear scenarios, FLMFR, PFR and LSC methods fail as expected. Among the nonlinear methods and excluding FAMM, DISC seems to be the winner in the nonlinear smooth scenario but fails in the non--smooth one where only the FSAMFR and FKAMFR achieve similar values to nominal in the estimation set. In any case, there is an important loss in the prediction level. The extremely negative values in nonlinear non-smooth case are due to the smoothed predictions 
in the neighborhood of the discontinuities of the response (see Figure~\ref{fig:scen4}) from  FAMM, LSC and DISC methods.

\section{Real Data Applications}\label{sec:realapp}
In this section, we consider two real data applications to examine the performance of all competitor methods. A third example is available in the supplementary material. 

\subsection{Bike-sharing Data}

\begin{figure}
\centering
\includegraphics[scale=.45]{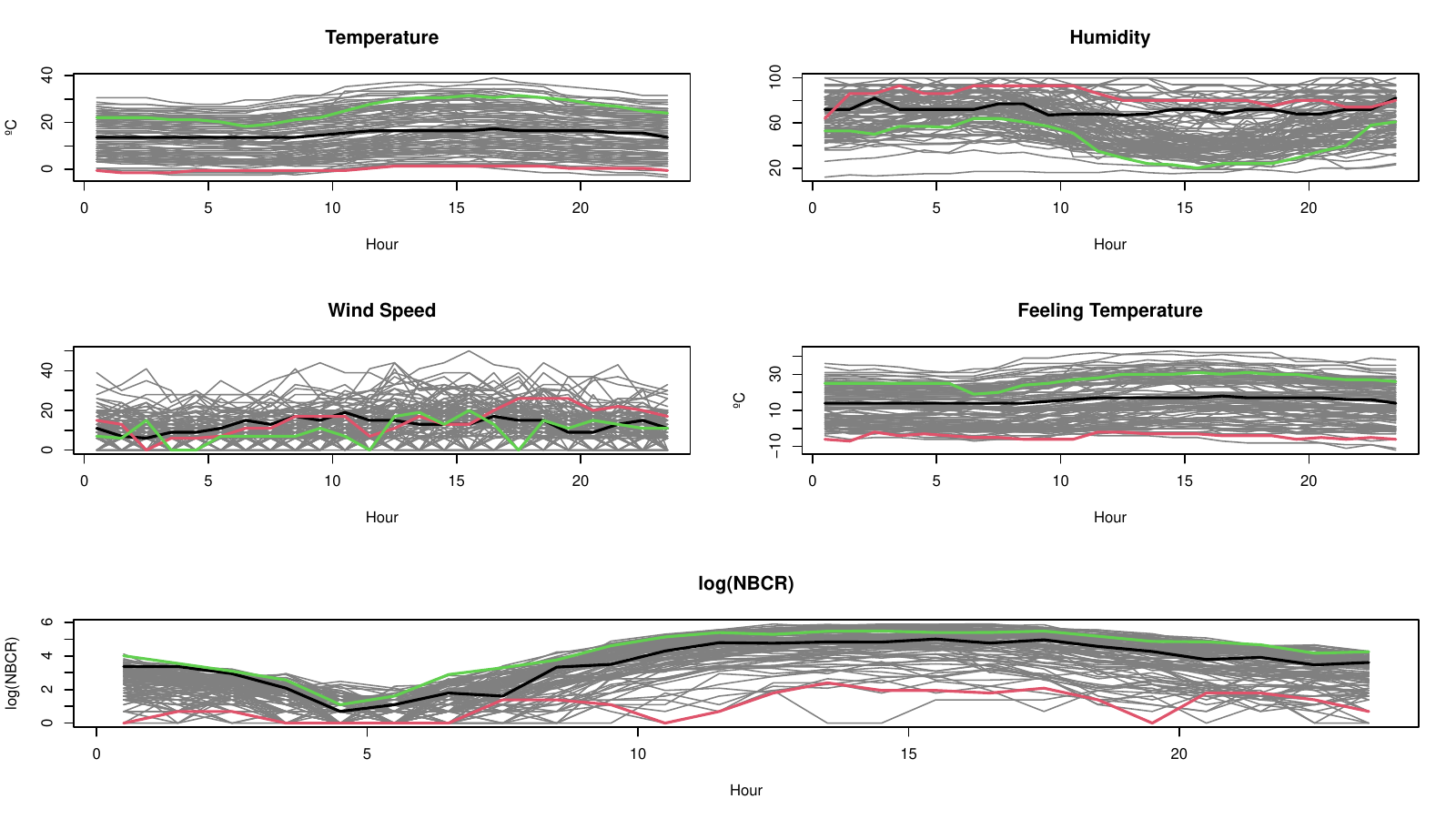}
\caption{The plot of Bike-sharing variables over 112 Saturdays from January \nth{1}, 2011 to December \nth{1}, 2012. The black, red, and green lines correspond to dates September  \nth{17}, 2011, January \nth{21}, 2012, and January \nth{9}, 2012, respectively.}
\label{bike_sh}
\end{figure}

To illustrate how our proposed function-on-function methods work, we use the Bike-sharing data (\cite{Fanaee-T2014}) as our first example. This dataset is collected by \href{https://ride.capitalbikeshare.com/system-data}{Capital Bikeshare System (CBS), Washington D.C., USA}.
The number of casual bike rentals (\textbf{NCBR}) is considered as our functional response, and Temperature (\textbf{T}), Humidity (\textbf{H}), Wind Speed (\textbf{WS}) and Feeling Temperature (\textbf{FT}), are the functional covariates. These variables are recorded each hour from January \nth{1}, 2011, to December \nth{1}, 2012. Similar to \cite{Kim2018}, we only consider the data for Saturday trajectories, and \textbf{NBCR} is log--transformed to avoid its natural heteroskedasticity. Ignoring three curves with missing values, the dataset contains 102 trajectories, each with 24 data points (hourly) for all variables. The corresponding plots are displayed in Figure \ref{bike_sh}. Table~\ref{DCbike} contains the distance correlation {(dCor)} between all variables (see \cite{Szekely2007}). It reveals that $\textbf{FT}$ has the highest distance correlation with the response ($\log(\textbf{NCBR})$).

\begin{table}[ht] 
\normalsize
\centering 
\begin{tabular}{cccccc} 
\hline\hline
\multicolumn{6}{c}{Distance correlation (dCor)}  \\ 
\hline
 & $\textbf{NCBR}$ & $\log(\textbf{NCBR}+1)$ &$\textbf{T}$ & $\textbf{H}$ & $\textbf{WS}$ \\ 
\hline 
$\textbf{T}$ & 0.527 & 0.678   & &   \\ 
$\textbf{H}$ &   0.067 & 0.069 & 0.054 &   \\ 
$\textbf{WS}$  &   0.062 & 0.102 & 0.057 & 0.053 \\
$\textbf{FT}$ & 0.552 & 0.705 & 0.994 & 0.054 & 0.060 \\
\hline 
\end{tabular}
\caption{Distance correlation among functional variables in Bike-sharing data.} 
\label{DCbike}  
\end{table}

Based on these values, a variable selection algorithm is performed similar to the one proposed  by \cite{FebreroBande2019}, and the covariates  \textbf{FT}, \textbf{H} and \textbf{WS} were selected as relevant for the response (in that order). Note that the selection algorithm does not select \textbf{T} due to its collinearity/concurvity with \textbf{FT} (dCor:$0.994$). The observations are randomly split into {\rm train} and {\rm test} sets (82 observations are considered for the train set, and the remaining $20$ are allocated to the {\rm test} set). We repeat this procedure 20 times and compute the $R_p^2$ for each one. Table \ref{tab_NuCaBiRe} shows the results for the aforementioned methods once, including the submodels in order of their relevance. The median of $R^2_p$ is shown instead of the mean as FAMM fails to converge in some cases (2--4 repetitions) due to numerical instabilities.     

\begin{table}
\centering\small
\resizebox{\textwidth}{!}{
\begin{tabular}{cccccccc}
\hline\hline
 \multicolumn{8}{c}{$\mcal{Y}=\log(\textbf{NCBR}+1)$}\\   \hline 
 covariates & FLMFR & FSAMFR & FKAMFR & PFR & FAMM & LSC & DISC \\ 
  \hline
  $\textbf{FT}$ & 0.550 & \textbf{0.621} & 0.609 & 0.543 & 0.398 & 0.544 & 0.592 \\ 
  $\textbf{FT}$, $\textbf{H}$ & 0.502 & 0.608 & \textbf{0.636} & 0.512 & 0.510 & 0.486 & \textbf{0.636} \\ 
  $\textbf{FT}$, $\textbf{H}$, $\textbf{WS}$ & 0.522 & 0.560 & 0.644 & 0.507 & 0.561 & 0.500 & \textbf{0.652} \\ 
   \hline\hline
\end{tabular}
}
\caption{Median of $R^2_p$ for the methods (including the covariates in order of importance) for predicting $\log(\textbf{NCBR}+1)$. The highest values per row are printed in bold.}\label{tab_NuCaBiRe}
\end{table}

 Table \ref{tab_NuCaBiRe} reveals that, whether we include all the covariates or only a subset of them, nonlinear methods provide better prediction performance than linear methods. However, the differences between the two groups are slight. With \textbf{FT}, all methods achieve a certain level that is slightly modified when we add more covariates. The addition of covariates slightly improves the $R^2_p$ for FKAMFR and DISC. It also slightly disimproves the remaining methods. The behaviour of FAMM is an exception showing an increasing trend but begins with the lowest result for the model with one covariate. The slight improvement from one to two or from two to three covariates suggests that the inclusion of the third covariate (\textbf{WS}) adds more uncertainty to estimation than prediction power.
 
 An important issue that deserves a detailed discussion is the computational cost. In this example, when needed, \textbf{FT} is represented in two PCs, \textbf{H} in four PCs, and \textbf{WS} in six PCs to ensure at least 95\% of the variability of each covariate is explained. The response requires six PCs to explain 90\% of  the variability. These parameters are related to the complexity that can be derived from the trace of the hat matrix $H$ ($\hat{\mcal{Y}}=H\mcal{Y}$). For instance, in FLMFR, each covariate consumes degrees of freedom equal to the number of PCs employed. In the case of FSAMFR, the consumed degrees of freedom  is $\sum_{j=1}^{nPCs} k_j$ where $k_j$ is the complexity of the function associated with the $j^{\rm th}$ PC (by default up to eight). Specifically, in this example,  $k_j$ is limited to five to have fewer parameters than the number of data. The effect on the complexity of PFR, FAMM, LSC, and DISC due to the number of PC of the covariates is not apparent. The reason is that, internally, these methods represent the information on a B--spline basis. Therefore, it seems that the complexity is mainly related to the length of the B--spline basis plus some penalizing factors like in the linear methods, PFR, and LSC. In the nonlinear methods, FAMM and DISC, this effect cannot be deduced more simply, and the help of these packages does not provide any further information. The complexity of FKAMFR is derived from the bandwidth values associated with each covariate, and it does not depend on the PC representation of the covariates. Regarding computational time\footnote{All computational times were obtained using the package \texttt{microbenchmark}.}, FLMFR, FSAMFR, LSC, and DISC obtain times measured in milliseconds ($51.5$, $266.2$, $112.9$, $559.5$, respectively), with important differences among them. PFR extends to $5.47$ seconds, and FKAMFR needs $27.32$ seconds, mostly employed in computing the optimal bandwidths and on the cycles of the backfitting algorithm. Finally, FAMM extends its time to $489.7$ seconds without a clear reason. Recall that the use of nonlinear contributions with \texttt{sff} command is considered an experimental feature by its authors.
 
\subsection{Electricity Demand and Price Data}
\begin{figure}
\centering
\includegraphics[scale=.4]{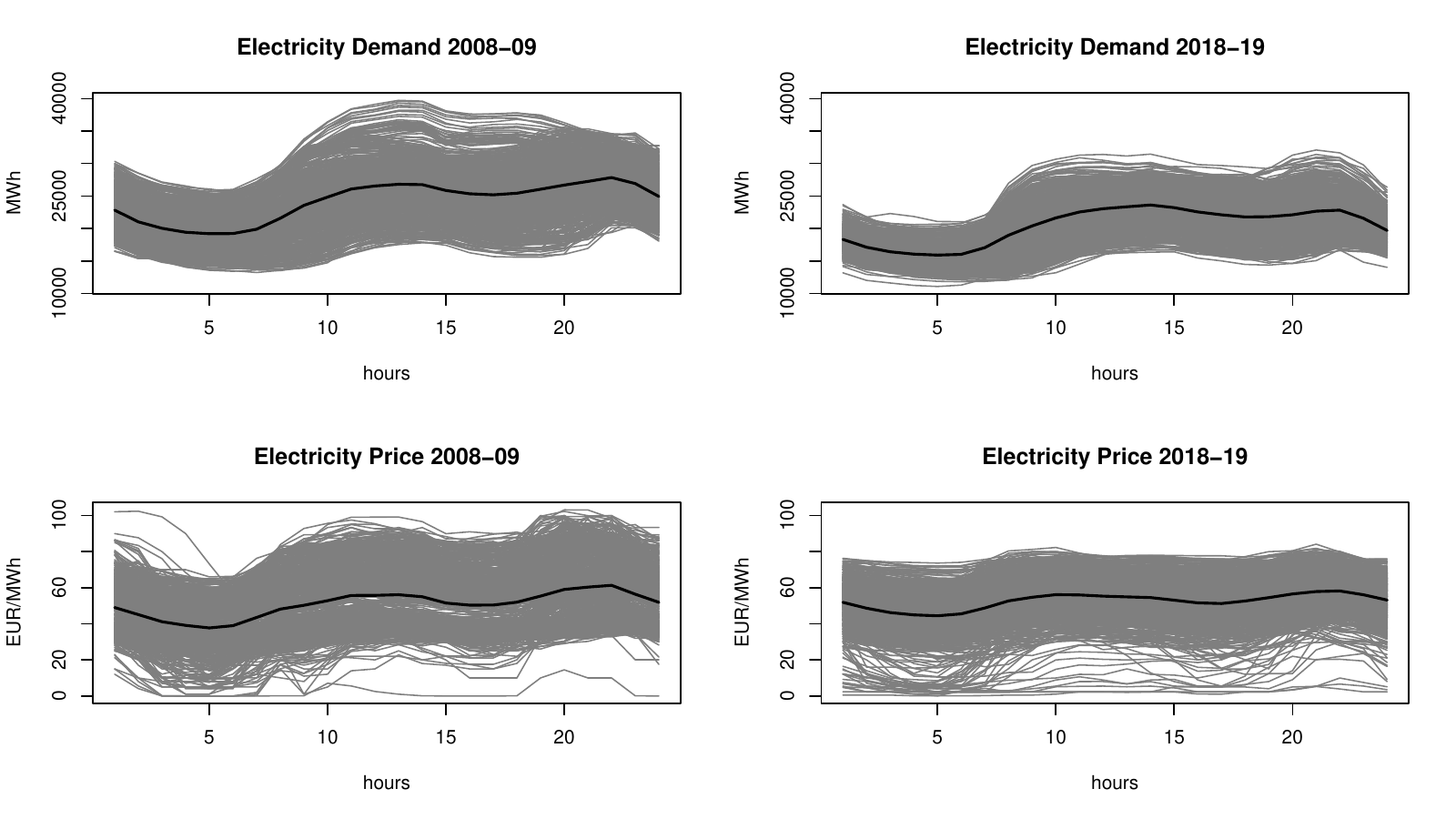}
\caption{Profiles for Electricity Demand (first row) and Electricity Price (second row) for the periods 2008-09 (first column) and 2018-19 (second column). The black line corresponds to the functional mean of each dataset.}
\label{omel}
\end{figure}

Another interesting example comes from the Iberian Energy Market. Specifically, the daily profiles of Electricity Price ($\textbf{Pr}_d$) and Demand ($\textbf{En}_d$) (the index refers to a day), both measured hourly, are obtained from two biannual periods separated by ten years: 2008-2009 and 2018-2019 (source:omie.es). \cite{FebreroBande2019} also employ this data, but they use a  different period and focus on variable selection for scalar responses. Comparing the two periods, the profiles (see Figure~\ref{omel}) show a considerable reduction in demand but not  a considerable price reduction. 

\begin{figure}
\centering
\includegraphics[scale=.4]{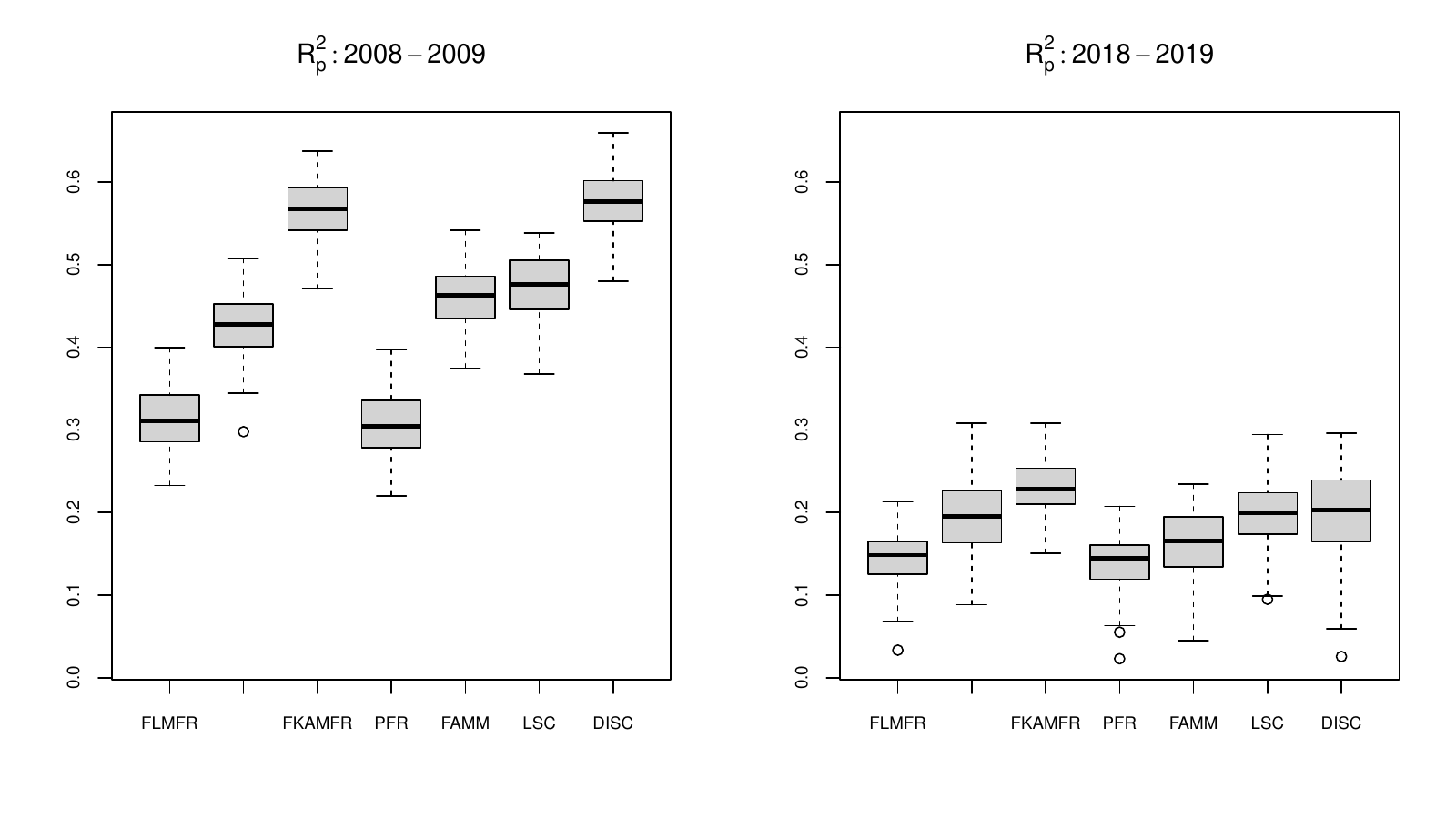}
\caption{Boxplots for $R^2_p$ ($100$ repetitions) for two different periods: 2008-2009 (left panel) and 2018-2019 (right panel).}
\label{omel2}
\end{figure}

In this example, a possible linear relationship among these two covariates is explored ($\textbf{Pr}_d$ explained by $\textbf{En}_d$) via a simulation study. In the simulation, we use $100$ repetitions, where each period is randomly split into a training and a prediction subset (75\%-25\%). All the models described above with only one covariate are applied to the training sample, and $R^2_p$ is computed from the prediction sample for impartial comparative purposes. When necessary, four principal components were employed for representing $\textbf{Pr}$ and $\textbf{En}$ (more than 95\% of explained variability is obtained in both cases). The results of the $100$ repetitions are shown in Table~\ref{tab:omel1} and Figure~\ref{omel2}, where a significant difference between the two periods is observed. The FKAMFR method obtains the best result in both cases, with the best result of $0.660$ for 2008-09. The other nonlinear methods obtain $0.576$ (DISC), $0.473$ (FAMM) and $0.428$ (FSAMFR). These differences between nonlinear methods suggest that the nonlinearity is due to some clusters owing to different patterns for groups of days (perhaps the differences among labour and weekend days). The linear methods obtain results between $0.473$ (LSC) and $0.306$ (PFR), suggesting that, in this example, there is a clear nonlinear relationship among $\textbf{Pr}_d$ and $\textbf{En}_d$. The conclusions for the second period are completely different. See the boxplots in Figure~\ref{omel2}, which are drawn at the same scale. The numerical results are between $0.262$ (FKAMFR) and $0.137$ (PFR), suggesting a weak linear relationship among covariates. As in this case, we have only one covariate. Therefore, we applied the test of linearity between a single functional covariate and a functional response described by \cite{GarciaPortugues2021}. For both periods, we obtain zero $p$-values and thus, we reject the linear hypothesis. Note that the test statistic for 2018-19 is relatively closer to the accepting region than 2008-09.

\begin{table}
\centering\small
\begin{tabular}{cccccccc}
\hline \hline
\multicolumn{8}{c}{$\mcal{Y}=\textbf{Pr}_d \sim \textbf{En}_d$}\\
\hline
 $\bar{R}_p^2$ & FLMFR & FSAMFR & FKAMFR & PFR & FAMM & LSC & DISC \\ 
  \hline
 2008-09 & 0.321 & 0.428 & \textbf{0.660} & 0.306 & 0.460 & 0.473 & 0.576 \\ \hline
 2018-19 & 0.143 & 0.194 & \textbf{0.262} & 0.137 & 0.161 & 0.197 & 0.201 \\  \hline
\end{tabular}
\caption{Mean of $R_p^2$ for the Energy data. Periods: 2008-09 and 2018-19.}\label{tab:omel1}
\end{table}

The relevant covariates selected in \cite{FebreroBande2019} to build predictive models (using information up to $d-1$ to explain $d$)  are mainly the lagged functional covariates at $d-1$ and $d-7$. Following the same idea, we implement a simpler numerical study ($10$ repetitions) to explain $\textbf{Pr}_d$ with  covariates $\textbf{En}_{d-1}$, $\textbf{En}_{d-7}$ and with $\textbf{Pr}_{d-1}$, $\textbf{Pr}_{d-7}$ using only the period 2008-09. Again, we split the data into training and  prediction samples (75\%-25\%) with $543$ curves in the training and $181$ in the prediction sample. Table~\ref{tab:omel3} summarizes the results. The first row (covariates  $\textbf{En}_{d-1}$, $\textbf{En}_{d-7}$) is similar to Table~\ref{tab:omel1} but with a rise of about $0.1$ for FAMM, LSC and DISC. These methods now being closer to FKAMFR and with a clear gap compared with linear methods. When the covariates are $\textbf{Pr}_{d-1}$ and $\textbf{Pr}_{d-7}$ (second row), the model seems linear because all models obtain almost the same high results (around $0.88$), FLMFR being the best.
This suggests that $\textbf{Pr}_d$ follows a persistent process, where the tomorrow's price is almost perfectly explained by today's and the last week price. In any case, the gap among the rows of Table~\ref{tab:omel3} suggests that there are other factors more important than the load/demand for determining the electricity prices.

Another important factor to take into account is the computational time. FLMFR and FSAMFR take about three seconds, while PFR, LSC, and DISC need 10 to 18 seconds. FAMM and FKAMFR go over two minutes to complete their computations. 

\begin{table}
\centering\small
\resizebox{\textwidth}{!}{
\begin{tabular}{cccccccc}
\hline\hline
\multicolumn{8}{c}{ $\mcal{Y}=\textbf{Pr}_d$ }\\
\hline 
$\bar{R}_p^2$   & FLMFR & FSAMFR & FKAMFR & PFR & FAMM & LSC & DISC \\ 
  \hline
 $\textbf{En}_{d-1}$, $\textbf{En}_{d-7}$ & 0.336 & 0.456 & 0.631 & 0.324 & 0.564 & 0.600 & \textbf{0.655} \\ \hline
 $\textbf{Pr}_{d-1}$, $\textbf{Pr}_{d-7}$ & \textbf{0.900} & 0.881 & 0.886 & 0.886 & 0.859 & 0.898 & 0.881 \\  \hline
\end{tabular}
}
\caption{Average of $R_p^2$ for prediction models with response $\textbf{Pr}_d$. Period: 2008-09.}\label{tab:omel3}
\end{table}

\section{Concluding Remarks}
In this paper, a review about functional regression models with functional response was provided jointly with two new nonlinear proposals: FSAMFR and FKAMFR. Both are extensions of functional regression models with scalar response and the main difference among them is the nature of the functional spaces associated with the response and the covariates. As FSAMFR is based on the representation of all functional information in bases, this proposal is restricted to be applied when the response and the covariates spaces are Hilbert. On the other hand, the estimation of a FKAMFR model mimics usual kernel regression models but using distances allowing to be applied to general metric or normed spaces. The new proposals have been implemented in both functions that will be available in next version of package \texttt{fda.usc} (the present implementation is provided in the suplementary material with the package \texttt{fda.usc.devel}). A new implementation of a functional linear regression model (FLMFR) is provided that, on the contrary to previous approaches, does not need to assume that the trajectories are smooth. The previous approaches, linear (PFR and LSC) or nonlinear (FAMM and DISC), always represent the estimates using a B--spline basis that can have difficulties in scenarios with discrete discontinuities (see, for instance, scenario 2 in the simulation study). These previous approaches are not similar respect to its availability. The performance of the methods included in \texttt{FRegSigCom} (LSC and DISC) is competitive but the development of the package seems to be discontinued and only old versions can be installed. The \texttt{refund} package (PFR and FAMM) has a continuous development and the performance of PFR seems reasonable when the trajectories are quite smooth. On the contrary, in our simulation scenarios, the application of FAMM, particularly with several functional covariates, was unable to obtain similar results than the other methods even though several combinations of the parameters have been tried.           

\backmatter

\paragraph{Supplementary material}
{The supplementary material of this paper} contains three appendices with codes, data and aside tools: 
\begin{enumerate}
	\item A visualization tool for Linear-Nonlinear regression among the functional response and a single functional covariate, both in Hilbert spaces.
	\item An application to Air Quality Data.
	\item Codes and Data needed for executing the simulation study and real data examples. This is also included in a GitHub repository \url{http://github.com/moviedo5/FRMFR/}.
\end{enumerate}

\paragraph{Acknowledgements}
The research by Manuel Febrero--Bande and Manuel Oviedo--de~la~Fuente has been partially supported by the Spanish Grant PID2020-116587GB-I00 funded by MCIN/AEI/10.13039/501100011033. 
Indeed, Manuel Oviedo--de~la~Fuente has been partially supported by Spanish Grants
PID2020-113578RB-I00 and PID2023-147127OB-I00 ``ERDF/EU'', also funded by MCIN/AEI/10.13039/501100011033/, by the Xunta de Galicia (Grupos de Referencia Competitiva ED431C-2024/14 and Centro de Investigacion del Sistema universitario de Galicia ED431G 2019/01), all of them through the European Regional Development Funds (ERDF), and by the EU through the FEDER Galicia 2021-27 operational program (Ref. ED431G 2023/01). The research by Mohammad Darbalaei and Morteza Amini is based upon research funded by Iran National Science Foundation (INSF) under project No. 99014748.
\bibliography{mainCS_2024R1-final.bib}


\begin{thebibliography}{42}
\ifx \bisbn   \undefined \def \bisbn  #1{ISBN #1}\fi
\ifx \binits  \undefined \def \binits#1{#1}\fi
\ifx \bauthor  \undefined \def \bauthor#1{#1}\fi
\ifx \batitle  \undefined \def \batitle#1{#1}\fi
\ifx \bjtitle  \undefined \def \bjtitle#1{#1}\fi
\ifx \bvolume  \undefined \def \bvolume#1{\textbf{#1}}\fi
\ifx \byear  \undefined \def \byear#1{#1}\fi
\ifx \bissue  \undefined \def \bissue#1{#1}\fi
\ifx \bfpage  \undefined \def \bfpage#1{#1}\fi
\ifx \blpage  \undefined \def \blpage #1{#1}\fi
\ifx \burl  \undefined \def \burl#1{\textsf{#1}}\fi
\ifx \doiurl  \undefined \def \doiurl#1{\url{https://doi.org/#1}}\fi
\ifx \betal  \undefined \def \betal{\textit{et al.}}\fi
\ifx \binstitute  \undefined \def \binstitute#1{#1}\fi
\ifx \binstitutionaled  \undefined \def \binstitutionaled#1{#1}\fi
\ifx \bctitle  \undefined \def \bctitle#1{#1}\fi
\ifx \beditor  \undefined \def \beditor#1{#1}\fi
\ifx \bpublisher  \undefined \def \bpublisher#1{#1}\fi
\ifx \bbtitle  \undefined \def \bbtitle#1{#1}\fi
\ifx \bedition  \undefined \def \bedition#1{#1}\fi
\ifx \bseriesno  \undefined \def \bseriesno#1{#1}\fi
\ifx \blocation  \undefined \def \blocation#1{#1}\fi
\ifx \bsertitle  \undefined \def \bsertitle#1{#1}\fi
\ifx \bsnm \undefined \def \bsnm#1{#1}\fi
\ifx \bsuffix \undefined \def \bsuffix#1{#1}\fi
\ifx \bparticle \undefined \def \bparticle#1{#1}\fi
\ifx \barticle \undefined \def \barticle#1{#1}\fi
\bibcommenthead
\ifx \bconfdate \undefined \def \bconfdate #1{#1}\fi
\ifx \botherref \undefined \def \botherref #1{#1}\fi
\ifx \url \undefined \def \url#1{\textsf{#1}}\fi
\ifx \bchapter \undefined \def \bchapter#1{#1}\fi
\ifx \bbook \undefined \def \bbook#1{#1}\fi
\ifx \bcomment \undefined \def \bcomment#1{#1}\fi
\ifx \oauthor \undefined \def \oauthor#1{#1}\fi
\ifx \citeauthoryear \undefined \def \citeauthoryear#1{#1}\fi
\ifx \endbibitem  \undefined \def \endbibitem {}\fi
\ifx \bconflocation  \undefined \def \bconflocation#1{#1}\fi
\ifx \arxivurl  \undefined \def \arxivurl#1{\textsf{#1}}\fi
\csname PreBibitemsHook\endcsname

\bibitem[\protect\citeauthoryear{Aneiros-P{\'e}rez and
  Vieu}{2006}]{Aneiros2006}
\begin{barticle}
\bauthor{\bsnm{Aneiros-P{\'e}rez}, \binits{G.}},
\bauthor{\bsnm{Vieu}, \binits{P.}}:
\batitle{Semi-functional partial linear regression}.
\bjtitle{Statistics \& Probability Letters}
\bvolume{76}(\bissue{11}),
\bfpage{1102}--\blpage{1110}
(\byear{2006})
\end{barticle}
\endbibitem

\bibitem[\protect\citeauthoryear{Buja et~al.}{1989}]{Buja1989}
\begin{barticle}
\bauthor{\bsnm{Buja}, \binits{A.}},
\bauthor{\bsnm{Hastie}, \binits{T.}},
\bauthor{\bsnm{Tibshirani}, \binits{R.}}:
\batitle{Linear smoothers and additive models}.
\bjtitle{Annals of Statistics}
\bvolume{17}(\bissue{2}),
\bfpage{453}--\blpage{510}
(\byear{1989})
\end{barticle}
\endbibitem

\bibitem[\protect\citeauthoryear{Beyaztas and Shang}{2020}]{Beyaztas2020}
\begin{barticle}
\bauthor{\bsnm{Beyaztas}, \binits{U.}},
\bauthor{\bsnm{Shang}, \binits{H.L.}}:
\batitle{On function-on-function regression: partial least squares approach}.
\bjtitle{Environmental and Ecological Statistics}
\bvolume{27}(\bissue{1}),
\bfpage{95}--\blpage{114}
(\byear{2020})
\doiurl{10.1007/s10651-019-00436-1}
\end{barticle}
\endbibitem

\bibitem[\protect\citeauthoryear{Cuevas et~al.}{2002}]{Cuevas2002}
\begin{barticle}
\bauthor{\bsnm{Cuevas}, \binits{A.}},
\bauthor{\bsnm{Febrero--Bande}, \binits{M.}},
\bauthor{\bsnm{Fraiman}, \binits{R.}}:
\batitle{Linear functional regression: the case of fixed design and functional
  response}.
\bjtitle{Canadian Journal of Statistics}
\bvolume{30}(\bissue{2}),
\bfpage{285}--\blpage{300}
(\byear{2002})
\end{barticle}
\endbibitem

\bibitem[\protect\citeauthoryear{Cardot et~al.}{1999}]{Cardot1999}
\begin{barticle}
\bauthor{\bsnm{Cardot}, \binits{H.}},
\bauthor{\bsnm{Ferraty}, \binits{F.}},
\bauthor{\bsnm{Sarda}, \binits{P.}}:
\batitle{Functional linear model}.
\bjtitle{Statistics \& Probability Letters}
\bvolume{45}(\bissue{1}),
\bfpage{11}--\blpage{22}
(\byear{1999})
\end{barticle}
\endbibitem

\bibitem[\protect\citeauthoryear{Cardot et~al.}{2003}]{Cardot2003}
\begin{barticle}
\bauthor{\bsnm{Cardot}, \binits{H.}},
\bauthor{\bsnm{Ferraty}, \binits{F.}},
\bauthor{\bsnm{Sarda}, \binits{P.}}:
\batitle{Spline estimators for the functional linear model}.
\bjtitle{Statistica Sinica}
\bvolume{13}(\bissue{3}),
\bfpage{571}--\blpage{592}
(\byear{2003})
\end{barticle}
\endbibitem

\bibitem[\protect\citeauthoryear{Chen et~al.}{2011}]{Chen2011}
\begin{barticle}
\bauthor{\bsnm{Chen}, \binits{D.}},
\bauthor{\bsnm{Hall}, \binits{P.}},
\bauthor{\bsnm{M{\"u}ller}, \binits{H.-G.}}:
\batitle{Single and multiple index functional regression models with
  nonparametric link}.
\bjtitle{Annals of Statistics}
\bvolume{39}(\bissue{3}),
\bfpage{1720}--\blpage{1747}
(\byear{2011})
\end{barticle}
\endbibitem

\bibitem[\protect\citeauthoryear{Cardot et~al.}{2007}]{Cardot2007a}
\begin{barticle}
\bauthor{\bsnm{Cardot}, \binits{H.}},
\bauthor{\bsnm{Mas}, \binits{A.}},
\bauthor{\bsnm{Sarda}, \binits{P.}}:
\batitle{{CLT} in functional linear regression models}.
\bjtitle{Probability Theory and Related Fields}
\bvolume{138}(\bissue{3}),
\bfpage{325}--\blpage{361}
(\byear{2007})
\end{barticle}
\endbibitem

\bibitem[\protect\citeauthoryear{Chiou et~al.}{2003}]{Chiou2003}
\begin{barticle}
\bauthor{\bsnm{Chiou}, \binits{J.-M.}},
\bauthor{\bsnm{M{\"u}ller}, \binits{H.-G.}},
\bauthor{\bsnm{Wang}, \binits{J.-L.}}:
\batitle{Functional quasi-likelihood regression models with smooth random
  effects}.
\bjtitle{Journal of the Royal Statistical Society: Series B (Statistical
  Methodology)}
\bvolume{65}(\bissue{2}),
\bfpage{405}--\blpage{423}
(\byear{2003})
\end{barticle}
\endbibitem

\bibitem[\protect\citeauthoryear{Chiou et~al.}{2016}]{Chiou2016}
\begin{barticle}
\bauthor{\bsnm{Chiou}, \binits{J.-M.}},
\bauthor{\bsnm{Yang}, \binits{Y.-F.}},
\bauthor{\bsnm{Chen}, \binits{Y.-T.}}:
\batitle{Multivariate functional linear regression and prediction}.
\bjtitle{Journal of Multivariate Analysis}
\bvolume{146},
\bfpage{301}--\blpage{312}
(\byear{2016})
\doiurl{10.1016/j.jmva.2015.10.003}
\end{barticle}
\endbibitem

\bibitem[\protect\citeauthoryear{Delaigle and Hall}{2012}]{Delaigle2012b}
\begin{barticle}
\bauthor{\bsnm{Delaigle}, \binits{A.}},
\bauthor{\bsnm{Hall}, \binits{P.}}:
\batitle{Methodology and theory for partial least squares applied to functional
  data}.
\bjtitle{Annals of Statistics}
\bvolume{40}(\bissue{1}),
\bfpage{322}--\blpage{352}
(\byear{2012})
\end{barticle}
\endbibitem

\bibitem[\protect\citeauthoryear{Febrero-Bande et~al.}{2017}]{Febrero2017}
\begin{barticle}
\bauthor{\bsnm{Febrero--Bande}, \binits{M.}},
\bauthor{\bsnm{Galeano}, \binits{P.}},
\bauthor{\bsnm{Gonz{\'a}lez--Manteiga}, \binits{W.}}:
\batitle{Functional principal component regression and functional partial
  least-squares regression: An overview and a comparative study}.
\bjtitle{International Statistical Review}
\bvolume{85}(\bissue{1}),
\bfpage{61}--\blpage{83}
(\byear{2017})
\end{barticle}
\endbibitem

\bibitem[\protect\citeauthoryear{Febrero-Bande and
  Gonz{\'a}lez-Manteiga}{2013}]{Febrero-Bande2013}
\begin{barticle}
\bauthor{\bsnm{Febrero--Bande}, \binits{M.}},
\bauthor{\bsnm{Gonz{\'a}lez--Manteiga}, \binits{W.}}:
\batitle{Generalized additive models for functional data}.
\bjtitle{TEST}
\bvolume{22}(\bissue{2}),
\bfpage{278}--\blpage{292}
(\byear{2013})
\doiurl{10.1007/s11749-012-0308-0}
\end{barticle}
\endbibitem

\bibitem[\protect\citeauthoryear{Febrero-Bande et~al.}{2019}]{FebreroBande2019}
\begin{barticle}
\bauthor{\bsnm{Febrero--Bande}, \binits{M.}},
\bauthor{\bsnm{Gonz{\'a}lez--Manteiga}, \binits{W.}},
\bauthor{\bsnm{{Oviedo--de la Fuente}}, \binits{M.}}:
\batitle{Variable selection in functional additive regression models}.
\bjtitle{Computational Statistics}
\bvolume{34}(\bissue{2}),
\bfpage{469}--\blpage{487}
(\byear{2019})
\end{barticle}
\endbibitem

\bibitem[\protect\citeauthoryear{Ferraty et~al.}{2012}]{Ferraty2012}
\begin{barticle}
\bauthor{\bsnm{Ferraty}, \binits{F.}},
\bauthor{\bsnm{Gonz{\'a}lez--Manteiga}, \binits{W.}},
\bauthor{\bsnm{Mart{\'i}nez--Calvo}, \binits{A.}},
\bauthor{\bsnm{Vieu}, \binits{P.}}:
\batitle{Presmoothing in functional linear regression}.
\bjtitle{Statistica Sinica}
\bvolume{22}(\bissue{1}),
\bfpage{69}--\blpage{94}
(\byear{2012})
\end{barticle}
\endbibitem

\bibitem[\protect\citeauthoryear{Ferraty et~al.}{2013}]{Ferraty2013}
\begin{barticle}
\bauthor{\bsnm{Ferraty}, \binits{F.}},
\bauthor{\bsnm{Goia}, \binits{A.}},
\bauthor{\bsnm{Salinelli}, \binits{E.}},
\bauthor{\bsnm{Vieu}, \binits{P.}}:
\batitle{Functional projection pursuit regression}.
\bjtitle{TEST}
\bvolume{22}(\bissue{2}),
\bfpage{293}--\blpage{320}
(\byear{2013})
\end{barticle}
\endbibitem

\bibitem[\protect\citeauthoryear{Fan and James}{2011}]{Fan2011}
\begin{botherref}
\oauthor{\bsnm{Fan}, \binits{Y.}},
\oauthor{\bsnm{James}, \binits{G.M.}}:
Functional additive regression.
Technical report,
Marshall School of Business, University of Southern California
(2011)
\end{botherref}
\endbibitem

\bibitem[\protect\citeauthoryear{Ferraty et~al.}{2012}]{Ferraty2012a}
\begin{barticle}
\bauthor{\bsnm{Ferraty}, \binits{F.}},
\bauthor{\bsnm{Keilegom}, \binits{I.V.}},
\bauthor{\bsnm{Vieu}, \binits{P.}}:
\batitle{Regression when both response and predictor are functions}.
\bjtitle{Journal of Multivariate Analysis}
\bvolume{109},
\bfpage{10}--\blpage{28}
(\byear{2012})
\doiurl{10.1016/j.jmva.2012.02.008}
\end{barticle}
\endbibitem

\bibitem[\protect\citeauthoryear{Fanaee-T and Gama}{2014}]{Fanaee-T2014}
\begin{barticle}
\bauthor{\bsnm{Fanaee-T}, \binits{H.}},
\bauthor{\bsnm{Gama}, \binits{J.}}:
\batitle{Event labeling combining ensemble detectors and background knowledge}.
\bjtitle{Progress in Artificial Intelligence}
\bvolume{2}(\bissue{2}),
\bfpage{113}--\blpage{127}
(\byear{2014})
\end{barticle}
\endbibitem

\bibitem[\protect\citeauthoryear{Ferraty and Vieu}{2006}]{Ferraty2006}
\begin{bbook}
\bauthor{\bsnm{Ferraty}, \binits{F.}},
\bauthor{\bsnm{Vieu}, \binits{P.}}:
\bbtitle{Nonparametric Functional Data Analysis: Theory and Practice}.
\bsertitle{Springer Series in Statistics}.
\bpublisher{Springer},
\blocation{New York, NY}
(\byear{2006})
\end{bbook}
\endbibitem

\bibitem[\protect\citeauthoryear{Ferraty and Vieu}{2009}]{Ferraty2009}
\begin{barticle}
\bauthor{\bsnm{Ferraty}, \binits{F.}},
\bauthor{\bsnm{Vieu}, \binits{P.}}:
\batitle{Additive prediction and boosting for functional data}.
\bjtitle{Computational Statistics \& Data Analysis}
\bvolume{53}(\bissue{4}),
\bfpage{1400}--\blpage{1413}
(\byear{2009})
\end{barticle}
\endbibitem

\bibitem[\protect\citeauthoryear{Goldsmith et~al.}{2011}]{Goldsmith2011}
\begin{barticle}
\bauthor{\bsnm{Goldsmith}, \binits{J.}},
\bauthor{\bsnm{Bobb}, \binits{J.}},
\bauthor{\bsnm{Crainiceanu}, \binits{C.-M.}},
\bauthor{\bsnm{Caffo}, \binits{B.}},
\bauthor{\bsnm{Reich}, \binits{D.}}:
\batitle{Penalized functional regression}.
\bjtitle{Journal of Computational and Graphical Statistics}
\bvolume{20}(\bissue{4}),
\bfpage{830}--\blpage{851}
(\byear{2011})
\doiurl{10.1198/jcgs.2010.10007}
\end{barticle}
\endbibitem

\bibitem[\protect\citeauthoryear{Garc\'{\i}a-Portugu\'{e}s
  et~al.}{2021}]{GarciaPortugues2021}
\begin{barticle}
\bauthor{\bsnm{Garc\'{\i}a-Portugu\'{e}s}, \binits{E.}},
\bauthor{\bsnm{\'{A}lvarez-Li\'{e}bana}, \binits{J.}},
\bauthor{\bsnm{\'{A}lvarez-P\'{e}rez}, \binits{G.}},
\bauthor{\bsnm{Gonz\'{a}lez-Manteiga}, \binits{W.}}:
\batitle{A goodness-of-fit test for the functional linear model with functional
  response}.
\bjtitle{Scandinavian Journal of Statistics. Theory and Applications}
\bvolume{48}(\bissue{2}),
\bfpage{502}--\blpage{528}
(\byear{2021})
\doiurl{10.1111/sjos.12486}
\end{barticle}
\endbibitem

\bibitem[\protect\citeauthoryear{Hall et~al.}{2006}]{Hall2006}
\begin{barticle}
\bauthor{\bsnm{Hall}, \binits{P.}},
\bauthor{\bsnm{M{\"u}ller}, \binits{H.G.}},
\bauthor{\bsnm{Wang}, \binits{J.L.}}:
\batitle{Properties of principal component methods for functional and
  longitudinal data analysis}.
\bjtitle{Annals of Statistics}
\bvolume{34}(\bissue{3}),
\bfpage{1493}--\blpage{1517}
(\byear{2006})
\end{barticle}
\endbibitem

\bibitem[\protect\citeauthoryear{Ivanescu et~al.}{2014}]{Ivanescu2014}
\begin{barticle}
\bauthor{\bsnm{Ivanescu}, \binits{A.E.}},
\bauthor{\bsnm{Staicu}, \binits{A.-M.}},
\bauthor{\bsnm{Scheipl}, \binits{F.}},
\bauthor{\bsnm{Greven}, \binits{S.}}:
\batitle{Penalized function-on-function regression}.
\bjtitle{Computational Statistics}
\bvolume{30}(\bissue{2}),
\bfpage{539}--\blpage{568}
(\byear{2014})
\doiurl{10.1007/s00180-014-0548-4}
\end{barticle}
\endbibitem

\bibitem[\protect\citeauthoryear{Jeon et~al.}{2021}]{Jeong2021}
\begin{barticle}
\bauthor{\bsnm{Jeon}, \binits{J.M.}},
\bauthor{\bsnm{Park}, \binits{B.U.}},
\bauthor{\bsnm{Van~Keilegom}, \binits{I.}}:
\batitle{{Additive regression for non-Euclidean responses and predictors}}.
\bjtitle{Annals of Statistics}
\bvolume{49}(\bissue{5}),
\bfpage{2611}--\blpage{2641}
(\byear{2021})
\doiurl{10.1214/21-AOS2048}
\end{barticle}
\endbibitem

\bibitem[\protect\citeauthoryear{James et~al.}{2009}]{James2009}
\begin{barticle}
\bauthor{\bsnm{James}, \binits{G.M.}},
\bauthor{\bsnm{Wang}, \binits{J.}},
\bauthor{\bsnm{Zhu}, \binits{J.}}:
\batitle{Functional linear regression that's interpretable}.
\bjtitle{Annals of Statistics}
\bvolume{37}(\bissue{5A}),
\bfpage{2083}--\blpage{2108}
(\byear{2009})
\end{barticle}
\endbibitem

\bibitem[\protect\citeauthoryear{Kim et~al.}{2018}]{Kim2018}
\begin{barticle}
\bauthor{\bsnm{Kim}, \binits{J.S.}},
\bauthor{\bsnm{Staicu}, \binits{A.-M.}},
\bauthor{\bsnm{Maity}, \binits{A.}},
\bauthor{\bsnm{Carroll}, \binits{R.J.}},
\bauthor{\bsnm{Ruppert}, \binits{D.}}:
\batitle{Additive function-on-function regression}.
\bjtitle{Journal of Computational and Graphical Statistics}
\bvolume{27}(\bissue{1}),
\bfpage{234}--\blpage{244}
(\byear{2018})
\end{barticle}
\endbibitem

\bibitem[\protect\citeauthoryear{Locantore et~al.}{1999}]{Locantore1999}
\begin{barticle}
\bauthor{\bsnm{Locantore}, \binits{N.}},
\bauthor{\bsnm{Marron}, \binits{J.}},
\bauthor{\bsnm{Simpson}, \binits{D.}},
\bauthor{\bsnm{Tripoli}, \binits{N.}},
\bauthor{\bsnm{Zhang}, \binits{J.}},
\bauthor{\bsnm{Cohen}, \binits{K.}},
\bauthor{\bsnm{Boente}, \binits{G.}},
\bauthor{\bsnm{Fraiman}, \binits{R.}},
\bauthor{\bsnm{Brumback}, \binits{B.}},
\bauthor{\bsnm{Croux}, \binits{C.}}:
\batitle{Robust principal component analysis for functional data}.
\bjtitle{TEST}
\bvolume{8}(\bissue{1}),
\bfpage{1}--\blpage{73}
(\byear{1999})
\end{barticle}
\endbibitem

\bibitem[\protect\citeauthoryear{Luo and Qi}{2017}]{Luo2017}
\begin{barticle}
\bauthor{\bsnm{Luo}, \binits{R.}},
\bauthor{\bsnm{Qi}, \binits{X.}}:
\batitle{Function-on-function linear regression by signal compression}.
\bjtitle{Journal of the American Statistical Association}
\bvolume{112}(\bissue{518}),
\bfpage{690}--\blpage{705}
(\byear{2017})
\doiurl{10.1080/01621459.2016.1164053}
\end{barticle}
\endbibitem

\bibitem[\protect\citeauthoryear{McLean et~al.}{2014}]{McLean2014}
\begin{barticle}
\bauthor{\bsnm{McLean}, \binits{M.W.}},
\bauthor{\bsnm{Hooker}, \binits{G.}},
\bauthor{\bsnm{Staicu}, \binits{A.-M.}},
\bauthor{\bsnm{Scheipl}, \binits{F.}},
\bauthor{\bsnm{Ruppert}, \binits{D.}}:
\batitle{Functional generalized additive models}.
\bjtitle{Journal of Computational and Graphical Statistics}
\bvolume{23}(\bissue{1}),
\bfpage{249}--\blpage{269}
(\byear{2014})
\doiurl{10.1080/10618600.2012.729985}
\end{barticle}
\endbibitem

\bibitem[\protect\citeauthoryear{M{\"u}ller and Yao}{2008}]{MullerYao2008}
\begin{barticle}
\bauthor{\bsnm{M{\"u}ller}, \binits{H.G.}},
\bauthor{\bsnm{Yao}, \binits{F.}}:
\batitle{Functional additive models}.
\bjtitle{Journal of the American Statistical Association}
\bvolume{103}(\bissue{484}),
\bfpage{1534}--\blpage{1544}
(\byear{2008})
\end{barticle}
\endbibitem

\bibitem[\protect\citeauthoryear{Preda and Saporta}{2005}]{Preda2005}
\begin{barticle}
\bauthor{\bsnm{Preda}, \binits{C.}},
\bauthor{\bsnm{Saporta}, \binits{G.}}:
\batitle{{PLS} regression on a stochastic process}.
\bjtitle{Computational Statistics \& Data Analysis}
\bvolume{48}(\bissue{1}),
\bfpage{149}--\blpage{158}
(\byear{2005})
\end{barticle}
\endbibitem

\bibitem[\protect\citeauthoryear{Preda and Schiltz}{2011}]{Preda2011}
\begin{bchapter}
\bauthor{\bsnm{Preda}, \binits{C.}},
\bauthor{\bsnm{Schiltz}, \binits{J.}}:
\bctitle{Functional {PLS} regression with functional response: the basis
  expansion approach}.
In: \bbtitle{Proceedings of the 14th Applied Stochastic Models and Data
  Analysis Conference},
pp. \bfpage{1126}--\blpage{1133}
(\byear{2011}).
\bcomment{Universit{\`a} di Roma La Spienza}
\end{bchapter}
\endbibitem

\bibitem[\protect\citeauthoryear{Qi and Luo}{2019}]{Qi2019}
\begin{barticle}
\bauthor{\bsnm{Qi}, \binits{X.}},
\bauthor{\bsnm{Luo}, \binits{R.}}:
\batitle{Nonlinear function-on-function additive model with multiple predictor
  curves}.
\bjtitle{Statistica Sinica}
\bvolume{29}(\bissue{2}),
\bfpage{719}--\blpage{739}
(\byear{2019})
\end{barticle}
\endbibitem

\bibitem[\protect\citeauthoryear{Reiss and Ogden}{2007}]{Reiss2007}
\begin{barticle}
\bauthor{\bsnm{Reiss}, \binits{P.T.}},
\bauthor{\bsnm{Ogden}, \binits{R.T.}}:
\batitle{Functional principal component regression and functional partial least
  squares}.
\bjtitle{Journal of the American Statistical Association}
\bvolume{102},
\bfpage{984}--\blpage{996}
(\byear{2007})
\end{barticle}
\endbibitem

\bibitem[\protect\citeauthoryear{Rao and Reimherr}{2023a}]{Rao2023b}
\begin{barticle}
\bauthor{\bsnm{Rao}, \binits{A.R.}},
\bauthor{\bsnm{Reimherr}, \binits{M.}}:
\batitle{Modern non-linear function-on-function regression}.
\bjtitle{Statistics and Computing}
\bvolume{33}(\bissue{6}),
\bfpage{130}
(\byear{2023})
\end{barticle}
\endbibitem

\bibitem[\protect\citeauthoryear{Rao and Reimherr}{2023b}]{Rao2023}
\begin{barticle}
\bauthor{\bsnm{Rao}, \binits{A.R.}},
\bauthor{\bsnm{Reimherr}, \binits{M.}}:
\batitle{Nonlinear functional modeling using {N}eural {N}etworks}.
\bjtitle{Journal of Computational and Graphical Statistics}
\bvolume{32}(\bissue{4}),
\bfpage{1248}--\blpage{1257}
(\byear{2023})
\doiurl{10.1080/10618600.2023.2165498}
\end{barticle}
\endbibitem

\bibitem[\protect\citeauthoryear{Ramsay and Silverman}{2005}]{Ramsay2005}
\begin{bbook}
\bauthor{\bsnm{Ramsay}, \binits{J.O.}},
\bauthor{\bsnm{Silverman}, \binits{B.W.}}:
\bbtitle{Functional Data Analysis}.
\bpublisher{Springer},
\blocation{New York, NY}
(\byear{2005})
\end{bbook}
\endbibitem

\bibitem[\protect\citeauthoryear{Sz{\'e}kely et~al.}{2007}]{Szekely2007}
\begin{barticle}
\bauthor{\bsnm{Sz{\'e}kely}, \binits{G.J.}},
\bauthor{\bsnm{Rizzo}, \binits{M.L.}},
\bauthor{\bsnm{Bakirov}, \binits{N.K.}}:
\batitle{Measuring and testing dependence by correlation of distances}.
\bjtitle{Annals of Statistics}
\bvolume{35}(\bissue{6}),
\bfpage{2769}--\blpage{2794}
(\byear{2007})
\end{barticle}
\endbibitem

\bibitem[\protect\citeauthoryear{Scheipl et~al.}{2015}]{Scheipl2015}
\begin{barticle}
\bauthor{\bsnm{Scheipl}, \binits{F.}},
\bauthor{\bsnm{Staicu}, \binits{A.-M.}},
\bauthor{\bsnm{Greven}, \binits{S.}}:
\batitle{Functional additive mixed models}.
\bjtitle{Journal of Computational and Graphical Statistics}
\bvolume{24}(\bissue{2}),
\bfpage{477}--\blpage{501}
(\byear{2015})
\doiurl{10.1080/10618600.2014.901914}
\end{barticle}
\endbibitem

\bibitem[\protect\citeauthoryear{Yao et~al.}{2005}]{Yao2005}
\begin{barticle}
\bauthor{\bsnm{Yao}, \binits{F.}},
\bauthor{\bsnm{M{\"u}ller}, \binits{H.-G.}},
\bauthor{\bsnm{Wang}, \binits{J.-L.}}:
\batitle{Functional data analysis for sparse longitudinal data}.
\bjtitle{Journal of the American Statistical Association}
\bvolume{100}(\bissue{470}),
\bfpage{577}--\blpage{590}
(\byear{2005})
\doiurl{10.1198/016214504000001745}
\end{barticle}
\endbibitem

\end{thebibliography}
\newpage

\vspace{2cm}
\begin{center}
\LARGE { Supplementary material}
\end{center}
\begin{center}
\LARGE { Functional Regression Models with Functional Response: A New Approach and a Comparative Study}
\end{center}
\vspace{0.25cm}
\begin{center}
{ Manuel Febrero--Bande, Manuel Oviedo--de la Fuente, Mohammad Darbalaei, and Morteza Amini 
}
\end{center}

\setcounter{figure}{0}
 \setcounter{table}{0}

\section*{A Visualization Tool about Linear-Nonlinear Relationships}
\begin{figure}
\centering
\includegraphics[scale=.3]{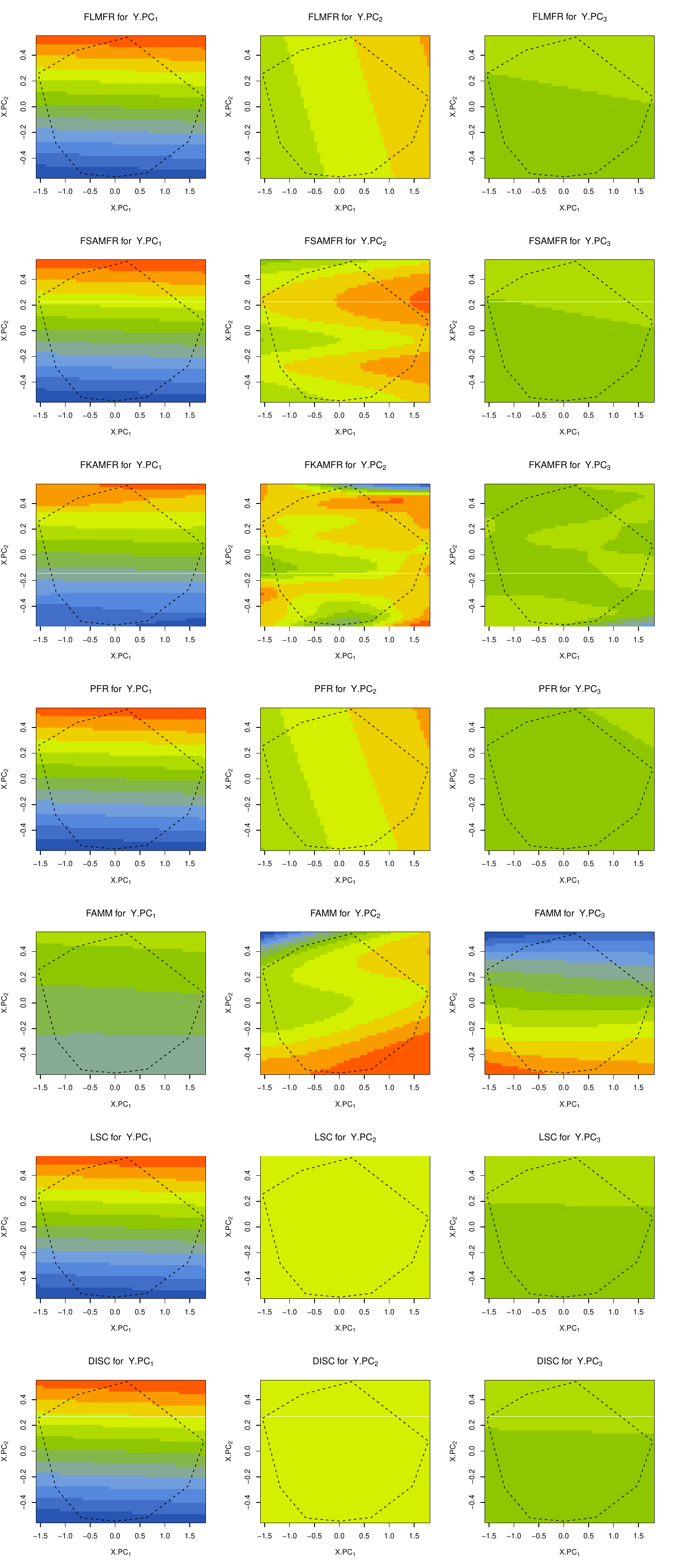}
\caption{Maps for predictive response PC scores (by columns) as a function  of the first two PC of the single covariate (Linear example). The dotted polygon inside each plot is the convex hull of the first two PCs of the covariate.}
\label{surfLin}
\end{figure}

\begin{figure}
\centering
\includegraphics[scale=.3]{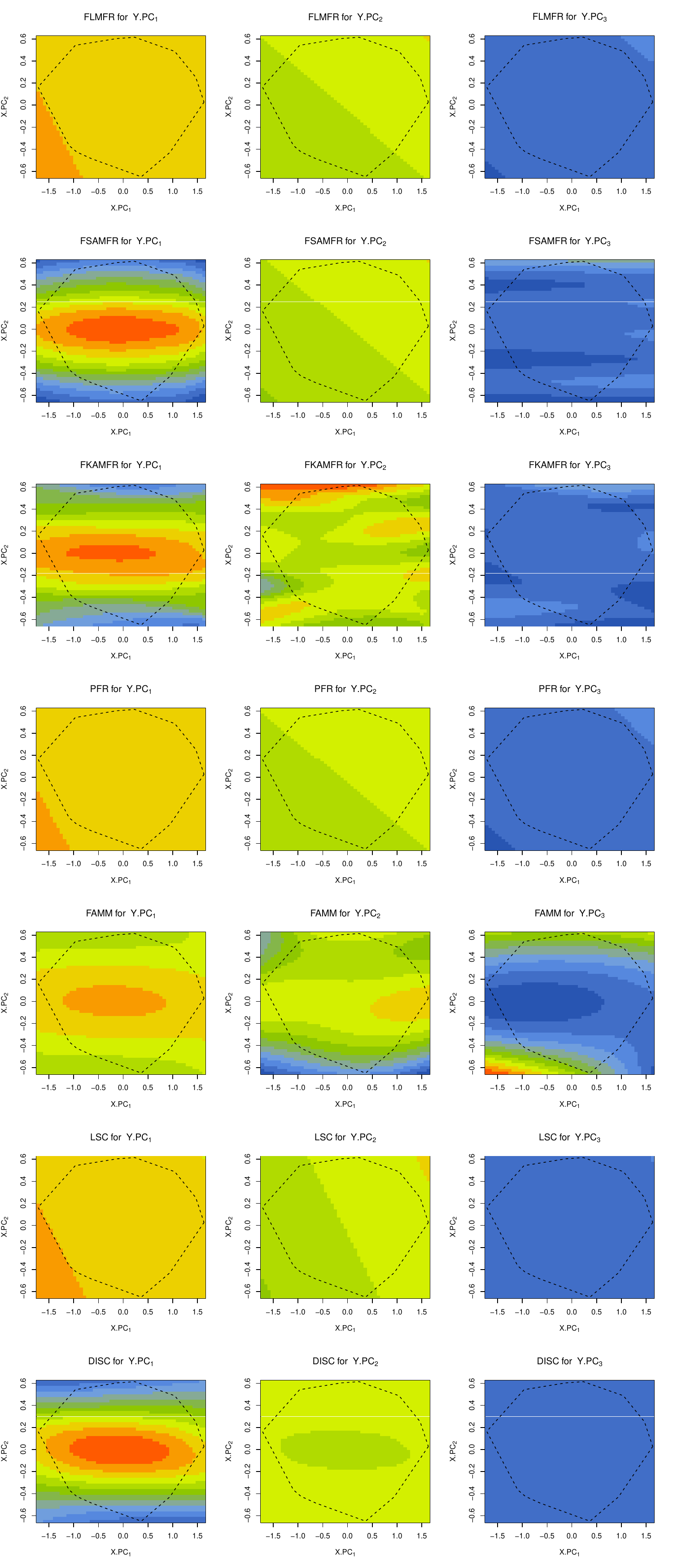}
\caption{Maps for predictive response PC scores (by columns) as a function  of the first two PC of the single covariate (Nonlinear example). The dotted polygon inside each plot is the convex hull of the first two PCs of the covariate.}
\label{surfNL}
\end{figure}

In multivariate framework, it could be relatively easy to construct some diagnostic plots for assessing when the partial contribution of a covariate is linear or nonlinear. This is extremely difficult in FDA due to the high dimensionality of objects involved. Nevertheless, something can be done in the case of Hilbertian response and a single Hilbertian covariate as shown in Figures~\ref{surfLin} and \ref{surfNL}. In both cases, we have a matrix of plots $7\times3$ where the columns corresponds with the $j^{\rm th}$ PC of the response ($Y.PC_j$).  The rows are associated with the models described in the paper. And, in each row-column combination, a map of the predicted score for the $j^{\rm th}$ of the response is plotted for a grid of predictor $\mcal{X}_0=\mu_{\mcal{X}}+x_{0,1}\eta_1+x_{0,2}\eta_2$ where $x_{0,1}$ and $x_{0,2}$ are generated in the range of the training sample and $\eta_1$, $\eta_2$ are the first two principal component of the covariate. All the plots have the same scale by columns in order to compare the different models. Also, the dotted line inside each map is the convex hull of  $x_{i,1}$ and $x_{i,2}$ from the training sample. The comparison among models must be done inside this polygon although the data cloud is not uniformly filling that convex hull.   

Figure~\ref{surfLin} corresponds to a linear example and the maps for all models are more or less the same except perhaps for FAMM (row 5) where a different pattern is shown (it seems smoother than the others in the first PC). The map for the first PC shows a clear decreasing trend from top to bottom that is also captured by the nonlinear methods (FSAMFR, FKAMFR and DISC). For the second PC, FSAMFR and FKAMFR provide a nonlinear pattern that seems not so different from the linear models inside the convex hull polygon.    

Figure~\ref{surfNL} shows an example of a nonlinear model that, as expected, is captured by FSAMFR, FKAMFR and DISC principally in the first PC (first column). FAMM captures a smoother version of the first PC and have a different pattern to the second and third PC (possibly to compensate its estimation of the first PC). The linear models (FLMFR, PFR and LSC) show a similar pattern among them different from the pattern captured by the nonlinear models.

For several covariates, a similar tool can be done fixing the other covariates to a particular values like, for instance, its mean and creating predictor vectors in the same way as before generating a grid using the two first PC components. But, it may happen that the plot seems linear because the nonlinearity were located far from the chosen fixed point. So, when this visualization tool shows a partial linear relationship, it is not a complete warranty about that assumption.

\section*{Air Quality Data}
Our final example is the Air Quality dataset (AQI) available from the UCI machine learning repository (\cite{Qi2019}). AQI is a popular dataset that consists of a list of five metal oxide chemical sensors embedded into an air quality multisensor device. The column names in the dataset begin with \texttt{PT}. The sensors are labelled as Carbon monoxide (\textbf{CO}), Non-methane hydrocarbons (\textbf{NMHC}), total Nitrogen Oxides (\textbf{NOx}), Ozone (\textbf{O3}) because it is supposed that its measures are related with the respective pollutants. The goal of this study is to predict the content of Benzene (\textbf{C6H6}) obtained through an independent analyzer which is considered the Ground Truth. The multisensor device is placed at a road level in an Italian city during a whole year. These signals were collected as a 24 hourly averaged concentration values in each day jointly with the relative humidity (\textbf{rH}) as external factors. 

\begin{figure}
\centering
\includegraphics[scale=.45]{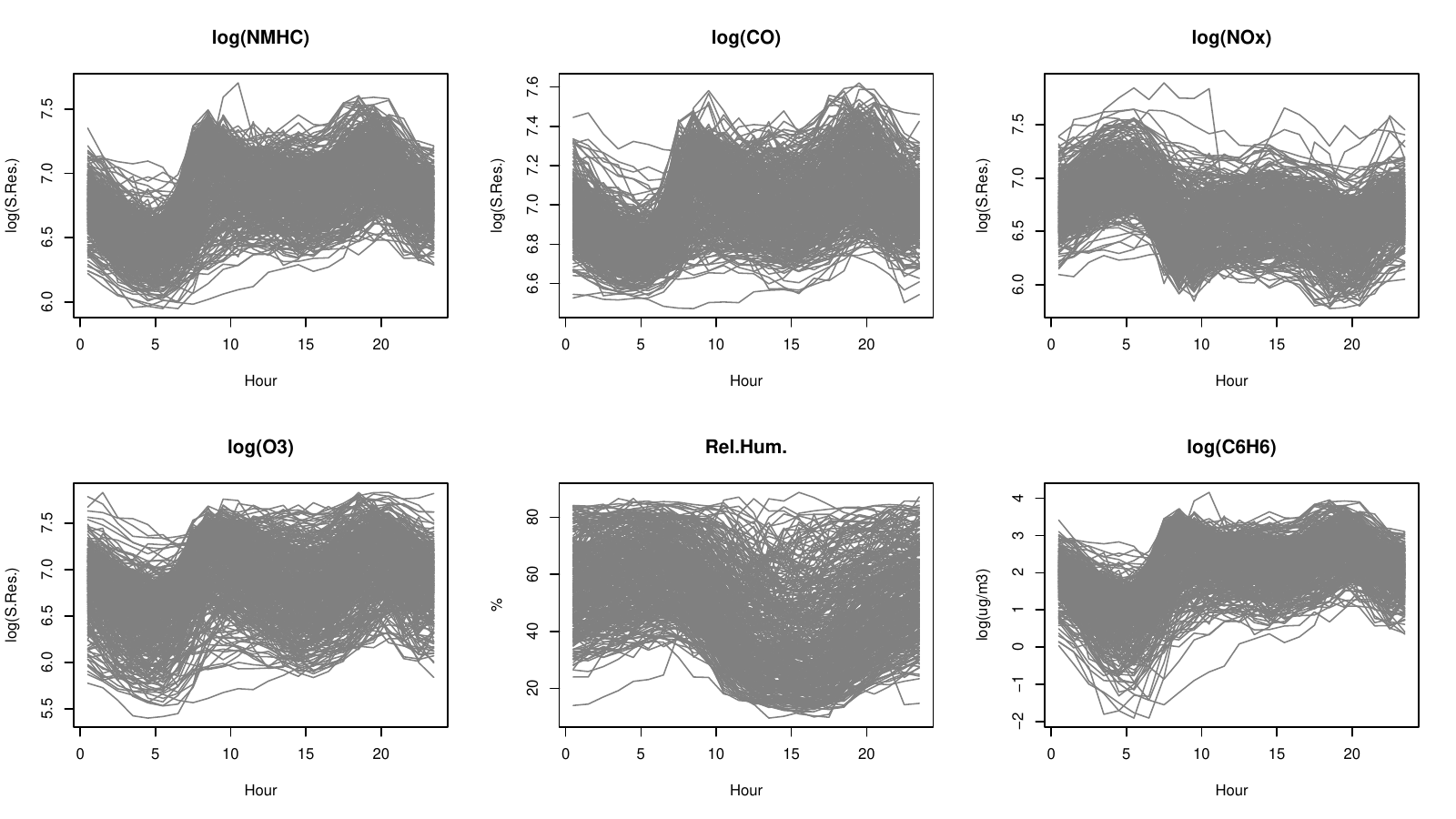}
\caption{355 curves for \textbf{NMHC, CO, NOx, O3} (obtained from sensors), and relative humidity jointly with the response \textbf{C6H6}. The signals from sensors are log--transformed.}
\label{AQII}
\end{figure}

After removing the curves with missing values, we have $355$ curves for each of the six functional variables shown in Figure \ref{AQII}.
Table \ref{DCaqi} presents the distance correlation between functional variables in the AQI dataset. We consider the curves of $\mathcal{Y}=\ln(\textbf{C6H6})$ as the functional response and the other five variables ($\mathcal{X}_1=\ln(\textbf{NMHC})$, $\mathcal{X}_2=\ln(\textbf{O3})$, $\mathcal{X}_3=\ln(\textbf{CO})$, $\mathcal{X}_4=\ln(\textbf{NOx})$ and $\mathcal{X}_5=\textbf{rH}$) as the functional {\rm pred}ictors.

\begin{table}[ht] 
\small
\centering 
\begin{tabular}{lccccc} 
\hline\hline
\multicolumn{6}{c}{Distance correlation}  \\ 
\hline
  & $\ln(\textbf{C6H6})$ &$\ln(\textbf{NMHC})$ & $\ln(\textbf{CO})$ & $\ln(\textbf{NOx})$ & $\ln(\textbf{O3})$ \\ 
\hline 
$\ln(\textbf{NMHC})$ & 0.986   &  &   &  &\\ 
$\ln(\textbf{CO})$ &   0.652 & 0.679  &  &  &\\ 
$\ln(\textbf{NOx})$  &   0.572 & 0.575 & 0.571 &  &\\
$\ln(\textbf{O3})$  &   0.791 & 0.800 & 0.782 & 0.683 & \\
$\textbf{rH}$ & 0.069 & 0.071 & 0.219 & 0.195 & 0.169 \\
\hline 
\end{tabular}
\caption{Distance correlation between functional  variables in the AQI dataset.} 
\label{DCaqi}  
\end{table}

In order to evaluate the performance of our proposed methods we randomly select 284 of the total observations as the {\rm train}ing set and the rest of 71 as the {\rm test} set. This procedure is repeated 100 times for the AQI dataset, and after {\rm fit}ting each method on the {\rm train}ing set, we estimate the regarding parameters for each method and then obtain the functional $R_p^2$  for each {\rm test} set. The averages of $R_p^2$ for the $100$ repetitions are tabulated in Table \ref{tab_C6H6}. In this table, sets M1, M2, and M3 consider different predictors. For M1, we select only one predictor, $\ln(\textbf{NMHC})$. For M2, we select three predictors: $\ln(\textbf{NMHC})$, $\ln(\textbf{O3})$, and $\ln(\textbf{CO})$. For M3, we use all the five predictors (namely, $\ln(\textbf{NMHC})$, $\ln(\textbf{O3})$, $\ln(\textbf{CO})$, $\ln(\textbf{NOx})$, and $\ln(\textbf{rH})$). Note that we select the variables in sets M1 and M2 similarly to \cite{FebreroBande2019}.

The results of Table \ref{tab_C6H6} reveals that the performance of linear methods are roughly the same as the nonlinear methods on all the three sets (M1, M2, M3). The simulation in the paper, also shows that for the inherently linear models the performance of linear and nonlinear methods is almost the same. Thus, it seems that the function-on-function model follows a linear behavior between response and covariates on AQI. In Table \ref{tab_C6H6}, the results of LSC method have always shown the highest performance. With a slight difference, DISC and FSAMFR methods have the second and third best performances. Moreover, Table \ref{tab_C6H6} reveals that the mentioned variable selection algorithm works well with this dataset. The reason is that in all the seven methods, increasing the number of model predictors, does not improve the performances considerably.

\begin{table}
\centering\small
\resizebox{\textwidth}{!}{
\begin{tabular}{lccccccc}
\hline\hline
 \multicolumn{8}{c}{$\mcal{Y}=\ln(\textbf{C6H6})$}\\   \hline 
 Model & FLMFR & FSAMFR & FKAMFR & PFR & FAMM & LSC & DISC \\ 
  \hline
  M1 & 0.889 & 0.900 & 0.848 & 0.855 & 0.815 & 0.923 & 0.920 \\ 
  M2 & 0.892 & 0.901 & 0.858 & 0.864 & 0.817 & 0.926 & 0.912 \\ 
  M3 & 0.891 & 0.900 & 0.859 & 0.862 & 0.815 & 0.924 & 0.899 \\ 
   \hline\hline
\end{tabular}
}
\caption{Average of $R^2_p$ over 100 procedures for the models (including the covariates in order of importance) for predicting $\ln(\textbf{C6H6})$. M1=$\{\mcal{X}_1\}$, M2=$\{\mcal{X}_1, \mcal{X}_2, \mcal{X}_3\}$, and M3=$\{\mcal{X}_1, \mcal{X}_2, \mcal{X}_3, \mcal{X}_4, \mcal{X}_5 \}.$
}
\label{tab_C6H6}
\end{table}

\section*{Supplementary Codes and Data}
\begin{itemize}
    \item \texttt{Simulation4CesgaS.R}: Code for main simulation. Scenarios 1--4.
    \item \texttt{bike-sharing2.R}: Code for Bike--sharing data example. 
    \item \texttt{hour.csv}: Bike--sharing data. 
    \item \texttt{Exampleomel.R}: Code for Electricity Demand and Price example.
    \item \texttt{omel2008-09.rda}: Electricity data for 2008--09 period.
    \item \texttt{omel2018-19.rda}: Electricity data for 2018--19 period.
    \item \texttt{data-real-aqi.R}: Code for example Air Quality.
    \item \texttt{AirQualityUCI.xlsx}: Air Quality Data.
    \item \texttt{fda.usc.devel\_2.0.4.tar.gz}: Package \texttt{fda.usc.devel} (Devel version of \texttt{fda.usc}) necessary for running the previous codes. 
\end{itemize}

\end{document}